\documentclass{jfp1}
\def\trimmarks{}
\expandafter\let\csname equation*\endcsname\relax
\expandafter\let\csname endequation*\endcsname\relax
\usepackage{amsmath}
\usepackage{paralist}
\usepackage{color}
\usepackage{url}
\usepackage[round,sort&compress]{natbib}
\usepackage{listings}

\usepackage{mathabx}

\usepackage{footmisc}
\usepackage[utf8]{inputenc}
\DeclareUnicodeCharacter{03BB}{\lambda} 
\DeclareUnicodeCharacter{FE52}{\;.\;} 
\DeclareUnicodeCharacter{03C0}{\pi} 
\DeclareUnicodeCharacter{211D}{\Re} 
\DeclareUnicodeCharacter{2192}{\rightarrow} 
\DeclareUnicodeCharacter{03B1}{\alpha} 
\DeclareUnicodeCharacter{2148}{\mathrm{i}} 
\DeclareUnicodeCharacter{1D53B}{\DD} 
\DeclareUnicodeCharacter{1D49F}{\D} 
\DeclareUnicodeCharacter{27F9}{\reducesto} 
\DeclareUnicodeCharacter{2080}{_0} 
\DeclareUnicodeCharacter{2081}{_1} 
\DeclareUnicodeCharacter{2082}{_2} 
\DeclareUnicodeCharacter{FF5C}{\mid} 
\DeclareUnicodeCharacter{03B5}{\ve} 
\DeclareUnicodeCharacter{2026}{\ldots} 
\DeclareUnicodeCharacter{22EF}{\cdots} 

\hypersetup{hidelinks,breaklinks,colorlinks,linkcolor={blue},citecolor={blue},urlcolor={blue}}
\renewcommand{\Re}{\mathbb{R}}
\newcommand{\ve}{\varepsilon}
\newcommand{\D}{{\mathcal D}}
\newcommand{\DD}{{\mathbb{D}}}
\newcommand{\J}{{\overrightarrow{{\mathcal J}}}}
\newcommand{\JJ}{{\overrightarrow{\mathbb{J}}}}
\newcommand{\Dh}{\hat{\mathcal D}}
\newcommand{\Dv}{\vec{\mathcal D}}
\DeclareMathOperator{\FIRST}{\mathbf{fst}}
\DeclareMathOperator{\SECOND}{\mathbf{snd}}
\DeclareMathOperator{\MAPPAIR}{\mathbf{mapPair}}
\DeclareMathOperator{\SQUARE}{\mathbf{sqr}}
\newcommand{\SUBST}[2]{[#1/#2]}
\DeclareMathOperator{\TANGENT}{\mathbf{tg}}
\DeclareMathOperator{\Tangent}{\mathbf{tangent}}
\DeclareMathOperator{\BUNDLE}{\mathbf{bun}}
\DeclareMathOperator{\Bundle}{\mathbf{bundle}}
\newcommand{\define}{\stackrel{\smalltriangleup}{=}} 
\DeclareMathOperator{\pf}{\mathbf{pf}}
\newcommand{\TANGENTSPACE}[1]{T_{#1}}
\newcommand{\tangent}[1]{#1'}
\newcommand{\fresh}{\mathbf{fresh}\;ε\;\mathbf{in}}
\newcommand{\LET}{\mathbf{let}}
\newcommand{\IN}{\mathbf{in}}
\newcommand{\reducesto}{\Longrightarrow} 
\newcommand{\equivalentto}{\Longleftrightarrow}
\DeclareSymbolFont{newfont}{OML}{cmm}{m}{it}
\DeclareMathSymbol{\nonvarepsilon}{3}{newfont}{15}
\newcommand{\eps}{\nonvarepsilon}
\newcommand{\justification}[1]{\hspace*{36pt}\{\text{#1}\}}

\makeatletter
\renewcommand\@biblabel[1]{}
\renewcommand\newblock{\hskip .11em \@plus .33em \@minus .07em}

\renewcommand\refname{References}
\renewenvironment{thebibliography}[1]
  {\section*{\refname}%
   \normalfont\small\rmfamily
   \addcontentsline{toc}{section}{\refname}%
   \list{}{\usecounter{enumiv}\labelwidth\z@ \leftmargin 9pt \itemindent -9pt}%
   \parindent\z@
   \parskip 2\p@ \@plus .1\p@
   \sloppy\clubpenalty\z@ \widowpenalty\@M
   \sfcode`\.\@m\relax}
  {\def\@noitemerr
   {\@latex@warning{Empty `thebibliography' environment}}%
   \endlist}
\makeatother

\newcommand{\eg}{\emph{e.g.},}
\newcommand{\Eg}{\emph{E.g.},}
\newcommand{\ie}{\emph{i.e.},}

\newcommand{\vs}{\emph{vs.}}

\newcommand{\defoccur}[1]{\textsl{#1}}

\newcommand{\Adifor}{\textsc{Adifor}}
\newcommand{\Tapenade}{\textsc{Tapenade}}
\newcommand{\FADBADplusplus}{\textsc{fadbad}$++$}
\newcommand{\SCMUTILS}{\textsc{scmutils}}
\newcommand{\HIPSAutograd}{\textsc{hips Autograd}}
\newcommand{\TorchAutograd}{\textsc{Torch Autograd}}
\newcommand{\Haskell}{\textsc{Haskell}}
\newcommand{\HaskellAD}{\textsc{Haskell ad}}

\newcommand{\RsixRSAD}{\textsc{r6rs-ad}}
\newcommand{\DiffSharp}{\textsc{DiffSharp}}
\newcommand{\Fsharp}{\textsc{f}$\sharp$}
\newcommand{\Julia}{\textsc{Julia}}
\newcommand{\Scheme}{\textsc{Scheme}}
\newcommand{\ML}{\textsc{ml}}
\newcommand{\Python}{\textsc{Python}}
\newcommand{\Lua}{\textsc{Lua}}
\newcommand{\Fortran}{\textsc{Fortran}}
\newcommand{\Clang}{\textsc{c}}
\newcommand{\Cplusplus}{\textsc{c}$++$}
\newcommand{\Myia}{\textsc{Myia}}
\newcommand{\TANGENTwilt}{\textsc{Tangent}}

\definecolor{darkred}{rgb}{0.7,0,0}
\definecolor{darkblue}{rgb}{0,0,0.7}

\title[Perturbation Confusion in Forward AD of Higher-Order Functions]
  {Perturbation Confusion in Forward Automatic
  Differentiation of Higher-Order Functions}

\author[O. Manzyuk et al.]
       {OLEKSANDR MANZYUK\thanks{Current affiliation: Facebook},
        BARAK A.\ PEARLMUTTER\thanks{
This work was supported, in part, by Science Foundation Ireland (SFI)
Principal Investigator grant 09/IN.1/I2637.},
        ALEXEY~ANDREYEVICH~RADUL\thanks{Current affiliation: Google AI}, and
        DAVID R.\ RUSH\thanks{Current address: Dunlavin, Ireland}\\
        Hamilton Institute \& Department of Computer Science\\
        Maynooth University, Co.~Kildare, Ireland\\
        \and\\
        JEFFREY MARK SISKIND\thanks{
          This work was supported, in part, by the Army Research Laboratory
          (ARL), accomplished under Cooperative Agreement Number
          W911NF-10-2-0060, by the National Science Foundation (NSF) under
          Grants 1522954-IIS and 1734938-IIS, and by the Intelligence Advanced
          Research Projects Activity (IARPA) via Department of
          Interior/Interior Business Center (DOI/IBC) contract number
          D17PC00341.
          Any opinions, findings, views, and conclusions or recommendations
          expressed in this material are those of the authors and do not
          necessarily reflect the views, official policies, or endorsements,
          either expressed or implied, of the SFI, ARL, NSF, IARPA, DOI/IBC,
          or the Irish or U.S.\ Governments.
          The U.S. Government is authorized to reproduce and distribute
          reprints for Government purposes, notwithstanding any copyright
          notation herein.}\\
        School of Electrical and Computer Engineering,\\
        Purdue University, West Lafayette IN 47907-2035, USA\\
        \email{\href{mailto:manzyuk@gmail.com}{manzyuk@gmail.com},
          \href{mailto:barak@pearlmutter.net}{barak@pearlmutter.net},
          \href{mailto:axch@alum.mit.edu}{axch@alum.mit.edu},
          \href{mailto:kumoyuki@gmail.com}{kumoyuki@gmail.com},
          \href{mailto:qobi@purdue.edu}{qobi@purdue.edu}}}

\begin{document}

\maketitle

\vspace*{-2ex}
\begin{abstract}
  \vspace*{-2ex}
  Automatic Differentiation (AD) is a technique for augmenting
  computer programs to compute derivatives.
  The essence of AD in its forward accumulation mode is to attach
  perturbations to each number, and propagate these through the computation by
  overloading the arithmetic operators.
  When derivatives are nested, the distinct derivative calculations, and
  their associated perturbations, must be distinguished.
  This is typically accomplished by creating a unique tag for each
  derivative calculation and tagging the perturbations.
  We exhibit a subtle bug, present in fielded implementations which support
  derivatives of higher-order functions, in which perturbations are confused
  \emph{despite} the tagging machinery, leading to incorrect results.
  The essence of the bug is this: a unique tag is needed for each derivative
  calculation, but in existing implementations unique tags are created when
  taking the derivative of a function at a point.
  When taking derivatives of higher-order functions, these need not correspond!
  We exhibit a simple example: a higher-order function $f$ whose derivative
  at a point $x$, namely $f'(x)$, is itself a function which calculates a
  derivative.
  This situation arises naturally when taking derivatives of curried functions.
  Two potential solutions are presented, and their deficiencies discussed.
  One uses eta expansion to delay the creation of fresh tags in order to put
  them into one-to-one correspondence with derivative calculations.
  The other wraps outputs of derivative operators with tag substitution
  machinery.
  Both solutions seem very difficult to implement without violating the
  desirable complexity guarantees of Forward AD.\@
  \vspace*{-2.5ex}
\end{abstract}

\section{Introduction}

The classical univariate derivative of a function $f:ℝ→ℝ$ is a
function $f':ℝ→ℝ$ \citep{Leibniz1664, Newton1704}.
Multivariate or vector calculus extends the notion of derivative to functions
whose domains and/or ranges are aggregates, \ie\ vectors, introducing notions
like gradients, Jacobians, and Hessians.
Differential geometry further extends the notion of derivatives to functions
whose domains and/or ranges are---or can contain---functions.

\defoccur{Automatic Differentiation} (AD) is a collection of methods for
computing the derivative of a function at a point when the function is
expressed as a computer program \citep{Griewank2008}.
These techniques, once pursued mainly by a small quiet academic
community, have recently moved to the forefront of deep learning,
where more expressive languages can spawn new industries, efficiency
improvements can save billions of dollars, and errors can have
far-reaching consequences.

From its earliest days, AD has supported functions whose domains and/or ranges
are aggregates.
There is currently interest from application programmers (machine learning in
particular) in applying AD to higher-order functions.
Here, we consider extending AD to support functions whose domains and/or
ranges are functions.
This is natural: we wish AD to be completely general and apply in an
unrestricted fashion to correctly compute the derivative of all programs that
compute differentiable mathematical functions.
This includes applying to functions whose domain and/or ranges include the
entire space of data types supported by programming languages, including not
only aggregates but also functions.
In doing so, we uncover a subtle bug.
Although for expository purposes we present the bug in the context of Forward
AD \citep{Wengert64}, the underlying issue can also manifest itself with other
AD modes, including Reverse AD \citep{Speelpenning1980} of higher-order
functions.
The bug is insidious: it can lead to production of incorrect results without
warning.
We present and discuss the relative merits of two fixes, and exhibit code
implementing them.

Our solutions are not ideal.
While we believe that the solutions will always produce the correct result,
they can foil both the space and time complexity guarantees of Forward AD
described in the next section.

\bigskip

Let~$𝔻$ denote the true mathematical derivative operator.
$𝔻$~is classically defined for first-order functions $ℝ→ℝ$ in
terms of limits and thus this classical definition does not lend itself to
direct implementation.
\begin{align}
  \label{eq:lim}
  𝔻\;f &= f' &
  \text{where
    \(f'(x)=\lim_{\nonvarepsilon→0}
    \frac{f(x+\nonvarepsilon)-f(x)}{\nonvarepsilon}\)}
\end{align}
We seek to materialize~$𝔻$ as a program construct~$𝒟$.
We can view this classical limit definition as a \emph{specification}
of~$𝒟$ and proceed to develop an \emph{implementation} of~$𝒟$.
Below, we use $=$ to denote mathematical equality, $\define$~to denote
definition of program constructs, and $⟹$ to denote evaluation.

One can extend~$𝔻$ to functions $ℝ→α$, where:
\begin{align}
  \label{eq:type}
  α::=ℝ｜α₁→α₂
\end{align}
We first focus on this extension in §\ref{sec:DA}--§\ref{sec:second}.
We consider further extension to functions
\mbox{$α₁→α₂$} in~§\ref{sec:DG}.
Since by~\eqref{eq:type} any type~$α$ must be of the form
$α₁→⋯→α_n→ℝ$, functions
$ℝ→α$ can be viewed as multivariate functions
$ℝ→α₂→⋯→α_n→ℝ$
whose first argument domain is~$ℝ$ and whose range is~$ℝ$.
We take $𝔻\;f$ where
\mbox{$f:ℝ→α₂→⋯→α_n→ℝ$}
to be the partial derivative with respect to the first argument.
\begin{align}
  \label{eq:partial}
  𝔻\;f=\frac{\partial f(x₁,x₂,…,x_n)}{\partial x₁}
\end{align}

We will see below that past work has implemented a~$𝒟$ that appears to
coincide with the specification~$𝔻$ in~\eqref{eq:lim} for
functions~$ℝ→ℝ$, but this past implementation fails to coincide
with the specification~$𝔻$ in~\eqref{eq:partial} for
functions~$ℝ→α$.
We then proceed to demonstrate two new implementations of~$𝒟$ that do appear to
coincide.

\section{Forward AD as Differential Algebra}
\label{sec:DA}

Forward AD can be formulated as differential algebra \citep{Karczmarczuk2001}.
Its essence is as follows.

The purely arithmetic theory of complex numbers as pairs of real numbers was
introduced by \citet{Hamilton1837}.
These form an algebra over two-term polynomials $a+bⅈ$ where
$ⅈ^2=-1$.
Arithmetic proceeds by simple rules, derived
algebraically.
\begin{subequations}
\begin{align}
  (a+bⅈ)+(c+dⅈ)&=(a+c)+(b+d)ⅈ\\
  (a+bⅈ)(c+dⅈ)&=ac+(ad+bc)ⅈ+bdⅈ^2=(ac-bd)+(ad+bc)ⅈ
\end{align}
\end{subequations}
Complex numbers can be implemented in a computer as ordered pairs $(a,b)$,
sometimes called Argand pairs.
Since arithmetic over complex numbers is defined in terms of arithmetic over
the reals, the above rules imply that computation over complex numbers is
closed.

\citet{Clifford1873} introduced \defoccur{dual numbers} of the form
$a+b\!\!\eps$.
In a dual number, the coefficient of~$\eps$ is called a perturbation or a
\defoccur{tangent}.
These can similarly be viewed as an algebra over two-term polynomials where
$\eps^2=0$ but $\eps\not=0$.
Arithmetic over dual numbers is again defined by simple rules derived
algebraically.
\begin{subequations}
\begin{align}
  \label{eq:plus}
  (a+b\!\!\eps)+(c+d\!\!\eps)&=(a+c)+(b+d)\!\!\eps\\
  \label{eq:times}
  (a+b\!\!\eps)(c+d\!\!\eps)&=ac+(ad+bc)\!\!\eps+bd\!\!\eps^2=ac+(ad+bc)\!\!\eps
\end{align}
\end{subequations}
Again, dual numbers can be implemented in a computer as ordered pairs $(a,b)$.
Again, since arithmetic over dual numbers is defined in terms of arithmetic
over the reals, the above rules imply that computation over dual numbers is
closed.

The essence of Forward AD is viewing dual numbers as truncated two-term power
series.
Since, following \citet{Taylor1715},
$f(x₀+x₁\!\!\eps+\,O(\eps^2))=f(x₀)+x₁f'(x₀)\!\!\eps+\,\,O(\eps^2)$, applying~$f$ to a dual number
$a+1\!\!\eps$ will yield a dual number $f(a)+f'(a)\!\!\eps$.
This leads to the following method for computing derivatives of functions
 $f:ℝ→ℝ$ expressed as computer programs.
\begin{compactitem}
\item Arrange for the programming language to support dual numbers and
  arithmetic thereupon.
\item To compute $f'$ at a point~$a$,
  \begin{compactenum}
  \item form $a+1\!\!\eps$,
    \label{step:A}
  \item apply~$f$ to $a+1\!\!\eps$ to obtain a result $f(a)+f'(a)\!\!\eps$,
    and
    \label{step:B}
  \item extract the tangent, $f'(a)$, from the result.
    \label{step:C}
  \end{compactenum}
\end{compactitem}
Step~\ref{step:B} constitutes a nonstandard interpretation of the arithmetic
basis functions with (\ref{eq:plus}, \ref{eq:times}).
This can be implemented in various ways, \eg\ overloading or source-code
transformation.
Further, dual numbers can be represented in various ways, \eg\ as unboxed
flattened values or as boxed values referenced through pointers.
These different implementation strategies do not concern us here.
While different implementation strategies have different costs, what we
discuss applies to all strategies.

It is convenient to encapsulate steps \ref{step:A}--\ref{step:C} as a
higher-order function $𝒟:f\mapsto f'$.
Indeed, that seems to be one of the original motivations for the development
of the lambda calculus \citep[¶4]{Church41}.
We can do this with the following code that implements~$𝒟$.
\begin{subequations}
\begin{align}
  \label{eq:A}
  \TANGENT\;a&\define 0&\text{$a : ℝ$}\\
  \label{eq:B}
  \TANGENT\;(a+b\!\!\eps)&\define b\\
  \label{eq:C}
  𝒟\;f\;x&\define\TANGENT\;(f\;(x+1\!\!\eps))
\end{align}
\end{subequations}
Here, $x+1\!\!\eps$ denotes step~\ref{step:A} above, \ie\ constructing a dual
number, and $\TANGENT\;(a+b\!\!\eps)$ denotes step~3 above, \ie\ extracting the
tangent of a dual number.
Equation~\eqref{eq:A} handles the case where the output of~$f$ is independent
of the input~$x$.

Forward AD provides certain complexity guarantees.
Steps~\ref{step:A} and~\ref{step:C} take unit time.
Step~\ref{step:B} introduces no more than a constant factor increase in both
the space and time complexity of executing~$f$ under a nonstandard
interpretation.
Thus computing $f\;x$ and $𝒟\;f\;x$ have the same space and time complexity.

\section{Tagging Dual Numbers to Avoid Perturbation Confusion}

\citet{SiskindPearlmutter2008a} discuss a problem with the above.
It is natural to nest application of~$𝒟$.
Doing so would allow taking higher-order derivatives and, more generally,
derivatives of functions that take derivatives of other functions.
\begin{align}
  \label{eq:Z}
  𝒟\;(λx﹒…𝒟\;(λy﹒…)\;…)\;…
\end{align}
This can lead to \defoccur{perturbation confusion} \citep[§2,
  Eqs.~4--11]{SiskindPearlmutter2005a}, yielding an incorrect result.
The essence of perturbation confusion is that each invocation of~$𝒟$ must
perform its computation over a distinct differential algebra.
While it is possible to reject programs that would exhibit perturbation
confusion using static typing \citep{Buckwalter2007, Kmett2010}, and static
typing can be used to yield the desired correct result in some cases with some
user annotation \citep{Shan2008}, no static method is known that can yield the
desired correct result in all cases without any annotation.
It is possible, however, to get the correct result in all cases (except, as we
shall see, when taking derivatives of functions whose ranges are functions)
without user annotation, by redefining $\TANGENT$ and $𝒟$ to \defoccur{tag}
dual numbers with distinct~$\eps$s to obtain distinct differential algebras
(or equivalently, distinct generators in a differential algebra) introduced by
different invocations of~$𝒟$ \citep{Lavendhomme-1996-sdg}.
We will indicate different tags by different subscripts on~$\eps$, and
use~$ε$ to denote a variable that is bound to an~$\eps$.
\begin{subequations}
\begin{align}
  \label{eq:D}
  \TANGENT\;ε\;a&\define 0&\text{$a:ℝ$}\\
  \label{eq:E}
  \TANGENT\;ε\;(a+bε)&\define b\\
  \label{eq:F}
  \TANGENT\;ε₁\;(a+bε₂)&\define
  (\TANGENT\;ε₁\;a)+(\TANGENT\;ε₁\;b)ε₂&ε₁\not=ε₂\\
  \label{eq:G}
  𝒟\;f\;x&\define\fresh\;\TANGENT\;ε\;(f\;(x+1ε))
\end{align}
These redefine (\ref{eq:A}--\ref{eq:C}).
Here, the tags are generated dynamically.
Many systems employ this
approach.\footnote{\label{foot:systemlist}\Eg\ \SCMUTILS\ \citep{Sussman1997a,
  Sussman1997b}, a software package that accompanies a textbook on classical
  mechanics \citep{SussmanWM2001} as well as a textbook on differential geometry
  \citep{sussman2013functional}, \citet{Farr2006},
  \citet{SiskindPearlmutter2005a, SiskindPearlmutter2008a},
  \citet{pearlmutter-siskind-popl-2007, Pearlmutter-Siskind-2008},
  \RsixRSAD\ (\url{https://github.com/qobi/R6RS-AD}),
  \DiffSharp\ \citep{baydin2016diffsharp},
  \HIPSAutograd\ \citep{maclaurin2015autograd},
  \TorchAutograd\ (\url{https://github.com/twitter/torch-autograd}), and
  \Julia\ (\url{http://www.juliadiff.org/ForwardDiff.jl/stable/user/api.html}).}
Many of these systems are implemented in `mostly functional languages,' like
\Scheme, \ML, \Fsharp, \Python, \Lua, and \Julia, and are intended to be used
with pure subsets of these languages.

Prior to this change, \ie\ with only a single~$\eps$, the values~$a$ and~$b$
in a dual number $a+b\!\!\eps$ would be real numbers.
With this change, \ie\ with multiple $\eps$s, the values~$a$ and~$b$ in a dual
number $a+b\!\!\eps₁$ can be dual numbers over~$\eps₂$ where
$\eps₂\neq\eps₁$.
Such a tree of dual numbers will contain real numbers in its leaves and will
contain a given~$\eps$ only once along each path from the root to the leaves.
Equation~\eqref{eq:F} provides the ability to extract the tangent of an~$\eps$
that might not be at the root of the tree.

\section{Extending to Functions whose Range is a Function}

If one applies~$𝒟$ to a function~$f$ whose range is a function, $f\;(x+1ε)$
in~\eqref{eq:G} will yield a function.
In this higher-order case, when~$f$ returns a function~$g$, an
invocation $𝒟\;f\;x$ yields a function~$\bar{g}$ which performs a derivative
calculation when invoked.
It will not be possible to extract the tangent of this with $\TANGENT$ as
implemented by (\ref{eq:D}--\ref{eq:F}).
The definition of $\TANGENT$ can be augmented to handle this case by
post-composition.\footnote{\label{foot:postcomposition}Justification of this
  post-composition is given in~§\ref{sec:DG} which describes the relevant
  constructs from differential geometry.}
\begin{align}
  \label{eq:I}
  \TANGENT\;ε\;\bar{g}&\define(\TANGENT\;ε)\circ\bar{g}&
  \text{$\bar{g}$ is a function}
\end{align}
\end{subequations}
However, this extension (alone) is flawed, as we proceed to demonstrate.

\section{A Bug}
\label{sec:bug}

Consider the following commonly occurring mathematical situation.
We define an offset operator:
\begin{gather}
  s:ℝ→(ℝ→ℝ)→ℝ→ℝ\nonumber\\
  \label{eq:s}
  s\;u\;f\;x\define f\;(x + u)
\end{gather}
The derivative of~$s$ at zero should be the same as the derivative
operator, \ie\ $𝔻\;s\;0=𝔻$, since:
\begin{subequations}
\begin{align}
  \label{eq:L1}
  (\forall f)(\forall y)𝔻\;s\;0\;f\;y
  &={\textstyle\frac{\partial}{\partial u}} [\:s\;u\; f\; y\:]_{u=0}
  ={\textstyle\frac{\partial}{\partial u}} [\:f(y+u)\:]_{u=0}
  =f'(y)
  =𝔻\;f\;y\\
  &\equivalentto\justification{eta}\nonumber\\
  \label{eq:L2}
  (\forall f)𝔻\;s\;0\;f&=𝔻\;f\\
  &\equivalentto\justification{eta}\nonumber\\
  \label{eq:L3}
  𝔻\;s\;0&=𝔻
\end{align}
\end{subequations}
Thus, if we define
\begin{equation}
  \label{eq:Dh-definition}
  \Dh\define𝒟\;s\;0
\end{equation}
we would hope that $\Dh=𝒟$.
However, we exhibit an example where it does not.

We can compute $\Dh\;(\Dh\;h)\;y$ for $h:ℝ→ℝ$ with simple
reduction steps:
\begin{subequations}
\begin{align}
  &\Dh\nonumber\\
  \label{eq:O1}
  ⟹&\justification{by~\eqref{eq:Dh-definition}}\nonumber\\
  &𝒟\;s\;0\\
  ⟹&\justification{by~\eqref{eq:G}}\nonumber\\
  \label{eq:O2}
  &\fresh\;\TANGENT\;ε\;(s\;(0+1ε))\\
  \label{eq:O3}
  ⟹&\justification{allocate a fresh tag~$\eps₀$; this is problematic; see
    discussion below}\nonumber\\
  &\TANGENT\;\eps₀\;(s\;(0+1\!\!\eps₀))\\
  \label{eq:O4}
  ⟹&\justification{by~\eqref{eq:s}}\nonumber\\
  &\TANGENT\;\eps₀\;(λf﹒λx﹒(f\;(x+1\!\!\eps₀)))\\
  \label{eq:O5}
  ⟹&\justification{by~\eqref{eq:I}}\nonumber\\
  &(\TANGENT\;\eps₀)\circ(λf﹒λx﹒(f\;(x+1\!\!\eps₀)))\\
  \label{eq:O6}
  ⟹&\justification{postcompose}\nonumber\\
  &λf﹒λx﹒\TANGENT\;\eps₀\;(f\;(x+1\!\!\eps₀))\\
  \hline
  \label{eq:O7}
  &\Dh\;(\Dh\;h)\;y\nonumber\\
  ⟹&\justification{substitute~\eqref{eq:O6} for~$\Dh$}\nonumber\\
  &(λf﹒λx﹒\TANGENT\;\eps₀\;(f\;(x+1\!\!\eps₀)))\;
   ((λf﹒λx﹒\TANGENT\;\eps₀\;(f\;(x+1\!\!\eps₀)))\;h)\;
   y\\
  \label{eq:O8}
  ⟹&\justification{beta reduce}\nonumber\\
  &(λf﹒λx﹒\TANGENT\;\eps₀\;(f\;(x+1\!\!\eps₀)))
  \;(λx﹒\TANGENT\;\eps₀\;(h\;(x+1\!\!\eps₀)))\;y\\
  \label{eq:O9}
  ⟹&\justification{beta reduce}\nonumber\\
  &(λx﹒\TANGENT\;\eps₀\;
  ((λx﹒\TANGENT\;\eps₀\;(h\;(x+1\!\!\eps₀)))\;(x+1\!\!\eps₀)))\;
  y\\
  \label{eq:O10}
  ⟹&\justification{beta reduce}\nonumber\\
  &\TANGENT\;\eps₀\;
  ((λx﹒\TANGENT\;\eps₀\;(h\;(x+1\!\!\eps₀)))\;(y+1\!\!\eps₀))\\
  \label{eq:O11}
  ⟹&\justification{beta reduce}\nonumber
\end{align}
\begin{align}
  &\TANGENT\;\eps₀\;(\TANGENT\;\eps₀\;(h\;((y+1\!\!\eps₀)+1\!\!\eps₀)))\\
  \label{eq:O12}
  ⟹&\justification{add dual numbers}\nonumber\\
  &\TANGENT\;\eps₀\;(\TANGENT\;\eps₀\;(h\;(y+2\!\!\eps₀)))\\
  \label{eq:O13}
  ⟹&\justification{apply~$h$ to a dual number}\nonumber\\
  &\TANGENT\;\eps₀\;(\TANGENT\;\eps₀\;(h(y)+2h'(y)\!\!\eps₀))\\
  \label{eq:O14}
  ⟹&\justification{by~\eqref{eq:E}}\nonumber\\
  &\TANGENT\;\eps₀\;(2h'(y))\\
  \label{eq:O15}
  ⟹&\justification{by~\eqref{eq:D}}\nonumber\\
  &0
\end{align}
\end{subequations}

This went wrong, yielding 0 instead of $h''(y)$.
\begin{equation}
  \label{eq:M}
  \Dh\;(\Dh\;h)\;y⟹0\not=𝔻\;(𝔻\;h)\;y=h''(y)
\end{equation}
The process of allocating a fresh tag in step~\eqref{eq:O4} was problematic.
The proper way to handle such fresh tag allocation might be to use nominal logic
\citep{pitts2003nominal}, perhaps in a dependent-type-theoretic variant
\citep{cheney2012dependent}.
Below, we offer alternate mechanisms that are suitable for use in
programming-language implementations that lack type systems that support first
class names and binding.

This is not an artificial example.
It is quite natural to construct an $x$-axis differential operator and apply
it to a two-dimensional function twice, along the~$x$ and then~$y$ axis
directions, by applying the operator, flipping the axes, and applying the
operator again, thus creating precisely this sort of cascaded use of a defined
differential operator.

\section{The Root Cause of the Bug}

This incorrect result was due to the tag~$\eps₀$ being generated exactly
\emph{once}, in~\eqref{eq:O2}, when $\Dh$ was calculated from $𝒟\;s\;0$ as
(\ref{eq:O1}--\ref{eq:O6}) using the definition~\eqref{eq:Dh-definition}.
The invocation $𝒟\;s\;0$ is the point at which a fresh tag is introduced;
early instantiation can result in reuse of the same tag in logically
distinct derivative calculations.
Here, the first derivative and the second derivative become confused
at~\eqref{eq:O12}.
We have two nested applications of $\TANGENT$ for~$\eps₀$, but for
correctness these should be distinctly tagged: $\eps₀$ \vs~$\eps₁$.

This can be accomplished by making two copies of~$\Dh$ by evaluating
$𝒟\;s\;0$ twice.
Performing an analogous computation with two copies of~$\Dh$ yields the correct
result.
\begin{subequations}
\begin{align}
  &\Dh₀\nonumber\\
  \label{eq:P1}
  ⟹&\justification{repeat~\eqref{eq:O1}}\nonumber\\
  &𝒟\;s\;0\\
  \label{eq:P2}
  ⟹&\justification{repeat~\eqref{eq:O2}}\nonumber\\
  &\fresh\;\TANGENT\;ε\;(s\;(0+1ε))\\
  \label{eq:P3}
  ⟹&\justification{repeat~\eqref{eq:O3}}\nonumber\\
  &\TANGENT\;\eps₀\;(s\;(0+1\!\!\eps₀))\\
  \label{eq:P4}
  ⟹&\justification{repeat~\eqref{eq:O4}}\nonumber
\end{align}
\begin{align}
  &\TANGENT\;\eps₀\;(λf﹒λx﹒(f\;(x+1\!\!\eps₀)))\\
  \label{eq:P5}
  ⟹&\justification{repeat~\eqref{eq:O5}}\nonumber\\
  &(\TANGENT\;\eps₀)\circ(λf﹒λx﹒(f\;(x+1\!\!\eps₀)))\\
  \label{eq:P6}
  ⟹&\justification{repeat~\eqref{eq:O6}}\nonumber\\
  &λf﹒λx﹒\TANGENT\;\eps₀\;(f\;(x+1\!\!\eps₀))\\
  \hline
  &\Dh₁\nonumber\\
  \label{eq:P7}
  ⟹&\justification{repeat~\eqref{eq:O1}}\nonumber\\
  &𝒟\;s\;0\\
  \label{eq:P8}
  ⟹&\justification{repeat~\eqref{eq:O2}}\nonumber\\
  &\fresh\;\TANGENT\;ε\;(s\;(0+1ε))\\
  \label{eq:P9}
  ⟹&\justification{repeat~\eqref{eq:O3}}\nonumber\\
  &\TANGENT\;\eps₁\;(s\;(0+1\!\!\eps₁))\\
  \label{eq:P10}
  ⟹&\justification{repeat~\eqref{eq:O4}}\nonumber\\
  &\TANGENT\;\eps₁\;(λf﹒λx﹒(f\;(x+1\!\!\eps₁)))\\
  \label{eq:P11}
  ⟹&\justification{repeat~\eqref{eq:O5}}\nonumber\\
  &(\TANGENT\;\eps₁)\circ(λf﹒λx﹒(f\;(x+1\!\!\eps₁)))\\
  \label{eq:P12}
  ⟹&\justification{repeat~\eqref{eq:O6}}\nonumber\\
  &λf﹒λx﹒\TANGENT\;\eps₁\;(f\;(x+1\!\!\eps₁))\\
  \hline
  &\Dh₀\;(\Dh₁\;h)\;y\nonumber\\
  \label{eq:P13}
  ⟹&\justification{substitute~\eqref{eq:P6} and~\eqref{eq:P12} for~$\Dh$}
  \nonumber\\
  &(λf﹒λx﹒\TANGENT\;\eps₀\;(f\;(x+1\!\!\eps₀)))\;
   ((λf﹒λx﹒\TANGENT\;\eps₁\;(f\;(x+1\!\!\eps₁)))\;h)\;
   y\\
  \label{eq:P14}
  ⟹&\justification{beta reduce}\nonumber\\
  &(λf﹒λx﹒\TANGENT\;\eps₀\;(f\;(x+1\!\!\eps₀)))
  \;(λx﹒\TANGENT\;\eps₁\;(h\;(x+1\!\!\eps₁)))\;y\\
  \label{eq:P15}
  ⟹&\justification{beta reduce}\nonumber\\
  &(λx﹒\TANGENT\;\eps₀\;
  ((λx﹒\TANGENT\;\eps₁\;(h\;(x+1\!\!\eps₁)))\;(x+1\!\!\eps₀)))\;
  y\\
  \label{eq:P16}
  ⟹&\justification{beta reduce}\nonumber\\
  &\TANGENT\;\eps₀\;
  ((λx﹒\TANGENT\;\eps₁\;(h\;(x+1\!\!\eps₁)))\;(y+1\!\!\eps₀))\\
  \label{eq:P17}
  ⟹&\justification{beta reduce}\nonumber\\
  &\TANGENT\;\eps₀\;(\TANGENT\;\eps₁\;(h\;((y+1\!\!\eps₀)+1\!\!\eps₁)))\\
  \label{eq:P18}
  ⟹&\justification{apply~$h$ to a dual number}\nonumber\\
  &\TANGENT\;\eps₀\;(\TANGENT\;\eps₁\;
  (h(y+1\!\!\eps₀)+h'(y+1\!\!\eps₀)\!\!\eps₁))\\
  \label{eq:P19}
  ⟹&\justification{apply~$h$ to a dual number}\nonumber\\
  &\TANGENT\;\eps₀\;(\TANGENT\;\eps₁\;
  ((h(y)+h'(y)\!\!\eps₀)+h'(y+1\!\!\eps₀)\!\!\eps₁))\\
  \label{eq:P20}
  ⟹&\justification{apply~$h$ to a dual number}\nonumber
\end{align}
\begin{align}
  &\TANGENT\;\eps₀\;(\TANGENT\;\eps₁\;
  ((h(y)+h'(y)\!\!\eps₀)+(h'(y)+h''(y)\!\!\eps₀)\!\!\eps₁))\\
  \label{eq:P21}
  ⟹&\justification{by~\eqref{eq:E}}\nonumber\\
  &\TANGENT\;\eps₀\;(h'(y)+h''(y)\!\!\eps₀)\\
  \label{eq:P22}
  ⟹&\justification{by~\eqref{eq:E}}\nonumber\\
  &h''(y)
\end{align}
\end{subequations}
Here, \eqref{eq:P18} corrects the mistake in~\eqref{eq:O12}.

However, this is tantamount to requiring the user to manually write
\begin{align}
  \begin{array}[t]{@{}l@{}}
    \LET\;\Dh₀\define𝒟\;s\;0\\
    \IN\;\begin{array}[t]{@{}l@{}}
    \LET\;\Dh₁\define𝒟\;s\;0\\
    \IN\;\Dh₀\;(\Dh₁\;h)\;y
    \end{array}
  \end{array}
\end{align}
instead of:
\begin{align}
  \begin{array}[t]{@{}l@{}}
    \LET\;\Dh\define𝒟\;s\;0\\
    \IN\;\Dh\;(\Dh\;h)\;y
  \end{array}
\end{align}
This should not be necessary since if~$𝒟$ correctly implemented~$𝔻$,
$\Dh₀$ and $\Dh₁$ should be equivalent.

The essence of the bug is that the implementation of~$𝒟$ in~\eqref{eq:G}
generates a distinct~$\eps$ for each invocation $𝒟\;f\;x$, but a
distinct~$\eps$ is needed for each derivative calculation.
In the first-order case, when $f:ℝ→ℝ$, these are equivalent.
Each invocation~$𝒟\;f\;x$ leads to a single derivative calculation.
But in the higher-order case, when~$f$ returns a function~$g$, an invocation
$𝒟\;f\;x$ yields~$\bar{g}$ which performs a derivative calculation when
invoked.
Since~$\bar{g}$ can be invoked multiple times, each such invocation will
perform a distinct derivative calculation and needs a distinct~$ε$.
The implementation in Appendix~\ref{app:implementation} illustrates the bug
when setting \lstinline{*eta-expansion?*} and \lstinline{*tag-substitution?*}
to \lstinline{#f} to use the definitions in~\eqref{eq:G} and~\eqref{eq:I}.

\section{A First Solution: Eta Expansion}
\label{sec:first}

One solution would be to eta expand the definition of~$𝒟$.
Such eta expansion would need to be conditional on the return type of~$f$.
\begin{subequations}
\begin{align}
   \label{eq:R}
  𝒟₁&:(ℝ→ℝ)→ℝ→ℝ\nonumber\\
  𝒟₁\;f\;x₁&\define\fresh\;\TANGENT\;ε\;(f\;(x₁+1ε))\\[2ex]
   \label{eq:S}
  𝒟₂&:(ℝ→α₂→ℝ)→ℝ→α₂
  →ℝ\nonumber\\
  𝒟₂\;f\;x₁\;x₂&\define\fresh\;\TANGENT\;ε\;(f\;(x₁+1ε)\;x₂)\\[2ex]
   \label{eq:T}
  𝒟_3&:(ℝ→α₂→α_3→ℝ)→ℝ
  →α₂→α_3→ℝ\nonumber\\
  𝒟_3\;f\;x₁\;x₂\;x_3&\define\fresh\;\TANGENT\;ε\;
  (f\;(x₁+1ε)\;x₂\;x_3)\\
  \vdots\nonumber
\end{align}
\end{subequations}
With such eta expansion conditioned on the return type of~$f$, \eqref{eq:I}
is not needed, because the appropriate variant of~$𝒟$ should only be invoked
in a context that contains all arguments necessary to subsequently allow the
call to $\TANGENT$ in that invocation of~$𝒟$ to yield to a
non-function-containing value.
This seemingly infinite set of~$𝒟_i$ and associated definitions can be
formulated as a single~$𝒟$ with polymorphic recursion.
\begin{subequations}
\begin{align}
  \label{eq:U}
  𝒟\;f\;x&\defineλy﹒(𝒟\;(λx﹒(f\;x\;y))\;x)&
  \text{$(f\;x)$ is a function}\\
  \label{eq:V}
  𝒟\;f\;x&\define\fresh\;\TANGENT\;ε\;(f\;(x+1ε))&
  \text{$(f\;x)$ is not a function}
\end{align}
\end{subequations}
We can see that this resolves the bug in (\ref{eq:O1}--\ref{eq:O15}) and
accomplishes the desiderata in (\ref{eq:P1}--\ref{eq:P12}) without making two
copies of~$\Dh$.
\begin{subequations}
  \begin{align}
  &\Dh\nonumber\\
  \label{eq:N1}
  ⟹&\justification{by~\eqref{eq:Dh-definition}}\nonumber\\
  &𝒟\;s\;0\\
  \label{eq:N2}
  ⟹&\justification{by~\eqref{eq:U}}\nonumber\\
  &λy﹒(𝒟\;(λx﹒(s\;x\;y))\;0)\\
  \hline
  &\Dh\;(\Dh\;h)\;y\nonumber\\
  \label{eq:N3}
  ⟹&\justification{substitute~\eqref{eq:N2} for~$\Dh$}\nonumber\\
  &
  (λy﹒(𝒟\;(λx﹒(s\;x\;y))\;0))
  \;((λy﹒(𝒟\;(λx﹒(s\;x\;y))\;0))\;h)\;y\\
  \label{eq:N4}
  ⟹&\justification{beta reduce}\nonumber\\
  &
  (λy﹒(𝒟\;(λx﹒(s\;x\;y))\;0))
  \;(𝒟\;(λx﹒(\;s\;x\;h))\;0)\;y\\
  \label{eq:N5}
  ⟹&\justification{beta reduce}\nonumber\\
  &
  (𝒟\;(λx﹒(s\;x\;(𝒟\;(λx﹒s\;x\;h)\;0)))\;0)\;y\\
  \label{eq:N6}
  ⟹&\justification{by~\eqref{eq:G}}\nonumber\\
  &
  (\fresh\;\TANGENT\;ε\;
  ((λx﹒(\;s\;x\;(𝒟\;(λx﹒(s\;x\;h))\;0)))\;(0+1ε)))\;
  y\\
  \label{eq:N7}
  ⟹&\justification{allocate a fresh tag~$\eps₀$}\nonumber\\
  &
  (\TANGENT\;\eps₀\;
  ((λx﹒(s\;x\;(𝒟\;(λx﹒(s\;x\;h))\;0)))\;
  (0+1\!\!\eps₀)))\;y\\
  \label{eq:N8}
  ⟹&\justification{beta reduce}\nonumber\\
  &
  (\TANGENT\;\eps₀\;(s\;(0+1\!\!\eps₀)\;(𝒟\;(λx﹒(s\;x\;h))\;0)))
  \;y\\
  \label{eq:N9}
  ⟹&\justification{by~\eqref{eq:G}}\nonumber\\
  &
  (\TANGENT\;\eps₀\;(s\;(0+1\!\!\eps₀)\;
  (\fresh\;\TANGENT\;ε\;((λx﹒(s\;x\;h))\;(0+1ε)))))\;y\\
  \label{eq:N10}
  ⟹&\justification{allocate a fresh tag~$\eps₁$}\nonumber\\
  &
  (\TANGENT\;\eps₀\;(s\;(0+1\!\!\eps₀)\;
  (\TANGENT\;\eps₁\;((λx﹒(s\;x\;h))\;(0+1\!\!\eps₁)))))\;y\\
  \label{eq:N11}
  ⟹&\justification{beta reduce}\nonumber\\
  &
  (\TANGENT\;\eps₀\;(s\;(0+1\!\!\eps₀)\;(\TANGENT\;\eps₁\;
  (s\;(0+1\!\!\eps₁)\;h))))\;y\\
  \label{eq:N12}
  ⟹&\justification{by~\eqref{eq:s}}\nonumber\\
  &
  (\TANGENT\;\eps₀\;(s\;(0+1\!\!\eps₀)\;
  (\TANGENT\;\eps₁\;(λx﹒(h\;(x+(0+1\!\!\eps₁)))))))\;y\\
  \label{eq:N13}
  ⟹&\justification{by~\eqref{eq:I}}\nonumber
\end{align}
\begin{align}
  &
  (\TANGENT\;\eps₀\;(s\;(0+1\!\!\eps₀)\;
  (\TANGENT\;\eps₁)\circ(λx﹒(h\;(x+(0+1\!\!\eps₁))))))\;y\\
  \label{eq:N14}
  ⟹&\justification{postcompose}\nonumber\\
  &(\TANGENT\;\eps₀\;(s\;(0+1\!\!\eps₀)\;
    (λx﹒(\TANGENT\;\eps₁\;(h\;(x+(0+1\!\!\eps₁)))))))\;y\\
  \label{eq:N15}
  ⟹&\justification{by~\eqref{eq:s}}\nonumber\\
  &(\TANGENT\;\eps₀\;(λx﹒
    ((λx﹒(\TANGENT\;\eps₁\;(h\;(x+(0+1\!\!\eps₁)))))\;
    (x+(0+1\!\!\eps₀)))))\;
    y\\
  \label{eq:N16}
  ⟹&\justification{beta reduce}\nonumber\\
  &
  (\TANGENT\;\eps₀\;(λx﹒
  (\TANGENT\;\eps₁\;(h\;((x+(0+1\!\!\eps₀))+(0+1\!\!\eps₁))))))\; y\\
  \label{eq:N17}
  ⟹&\justification{by~\eqref{eq:I}}\nonumber\\
  &
  (\TANGENT\;\eps₀)\circ(λx﹒
  (\TANGENT\;\eps₁\;(h\;((x+(0+1\!\!\eps₀))+(0+1\!\!\eps₁)))))\; y\\
  \label{eq:N18}
  ⟹&\justification{postcompose}\nonumber\\
  &
  (λx﹒(\TANGENT\;\eps₀\;
  (\TANGENT\;\eps₁\;(h\;((x+(0+1\!\!\eps₀))+(0+1\!\!\eps₁))))))\;y\\
  \label{eq:N19}
  ⟹&\justification{beta reduce}\nonumber\\
  &
  \TANGENT\;\eps₀\;
  (\TANGENT\;\eps₁\;(h\;((y+(0+1\!\!\eps₀))+(0+1\!\!\eps₁))))\\
  \label{eq:N20}
  ⟹&\justification{add dual numbers}\nonumber\\
  &
  \TANGENT\;\eps₀\;(\TANGENT\;\eps₁\;(h\;((y+1\!\!\eps₀)+(0+1\!\!\eps₁))))\\
  \label{eq:N21}
  ⟹&\justification{add dual numbers}\nonumber\\
  &
  \TANGENT\;\eps₀\;(\TANGENT\;\eps₁\;(h\;((y+1\!\!\eps₀)+1\!\!\eps₁)))\\
  \label{eq:N22}
  ⟹&\justification{same as~\eqref{eq:P18}}\nonumber\\
  &
  \TANGENT\;\eps₀\;
  (\TANGENT\;\eps₁\;(h(y+1\!\!\eps₀)+h'(y+1\!\!\eps₀)\!\!\eps₁))\\
  \label{eq:N23}
  ⟹&\justification{same as~\eqref{eq:P19}}\nonumber\\
  &\TANGENT\;\eps₀\;(\TANGENT\;\eps₁\;
  ((h(y)+h'(y)\!\!\eps₀)+h'(y+1\!\!\eps₀)\!\!\eps₁))\\
  \label{eq:N24}
  ⟹&\justification{same as~\eqref{eq:P20}}\nonumber\\
  &\TANGENT\;\eps₀\;(\TANGENT\;\eps₁\;
  ((h(y)+h'(y)\!\!\eps₀)+(h'(y)+h''(y)\!\!\eps₀)\!\!\eps₁))\\
  \label{eq:N25}
  ⟹&\justification{same as~\eqref{eq:P21}}\nonumber\\
  &\TANGENT\;\eps₀\;(h'(y)+h''(y)\!\!\eps₀)\\
  \label{eq:N26}
  ⟹&\justification{same as~\eqref{eq:P22}}\nonumber\\
  &h''(y)
\end{align}
\end{subequations}
Here, the allocation of a fresh tag is delayed from~\eqref{eq:N2} and is
performed twice, in~\eqref{eq:N7} and~\eqref{eq:N10}, allowing~\eqref{eq:N22}
to correct the mistake in~\eqref{eq:O12}, just like \eqref{eq:P18}.
The implementation in Appendix~\ref{app:implementation} illustrates that this
resolves the bug when setting \lstinline{*eta-expansion?*} to \lstinline{#t}
to use the definition in (\ref{eq:U}--\ref{eq:V}) instead of that in
\eqref{eq:G}.

\subsection{Issues with Eta Expansion}
\label{sec:first-issues}

This solution presents several problems.
\begin{itemize}
\item First, this manuscript only considers a space of types that includes
  scalar reals and functions but not aggregates (exclusive of dual numbers).
  Complications arise when extending the space of types to include aggregates.
  Appendix~\ref{app:implementation} illustrates that the above mechanism works
  with functions that return Church-encoded aggregates.
  \begin{subequations}
  \begin{align}
    (a,d)\;m&\define m\;a\;d\\
    \FIRST\;c&\define c\;(λa﹒(λd﹒a))\\
    \SECOND\;c&\define c\;(λa﹒(λd﹒d))\\
    t\;u&\define (e^{u\times u},(λf﹒(λx﹒(f\;x+u))))\\
    𝒟\;t\;1&⟹t'(1)\\
    p&\define𝒟\;t\;0\\
    \FIRST\;p&⟹0\\
    \Dv&\define\SECOND\;p\\
    \Dv\;(\Dv\;\exp)\;1&⟹e
  \end{align}
  \end{subequations}
  With a function that returned native aggregates, one would need to emulate
  the behavior that occurs with Church-encoded aggregates on native aggregates
  by delaying derivative calculation, with the associated tag allocation and
  $\TANGENT$ applied to the native returned aggregate, until an accessor is
  applied to that aggregate.
  Consider $𝒟\;t\;0$ where
  $t:ℝ→(ℝ\times((ℝ→ℝ)→ℝ))$ as above.
  One could not perform the derivative calculation when computing the
  value~$p$ returned by $𝒟\;t\;0$.
  One would have to delay until applying an accessor to~$p$.
  If one accessed the first element of~$p$, one would perform the derivative
  calculation, with the associated tag allocation, at the time of access.
  But if one accessed the second element of~$p$, one would have to further
  delay the derivative calculation, with the associated tag allocation, until
  that second element was invoked.
  This could require different amounts of delay that might be incompatible
  with some static type systems.
\item Second, with a type system or other static analysis mechanism that is
  unable to handle the unbounded polymorphism of (\ref{eq:R}, \ref{eq:S},
  \ref{eq:T}, …) or infer the ``is [not] a function'' side conditions of
  (\ref{eq:U}, \ref{eq:V}), achieving completeness might require run-time
  evaluation of the side conditions.
  This could involve calling~$f$ twice, once to determine its return type
  and once to do the eta-expanded derivative calculation, and lead to
  exponential increase in asymptotic time complexity.
\item Third, the solution can break sharing in curried functions, even with
  a type system or other static analysis mechanism that is able to eliminate
  the run-time evaluation of ``is [not] a function'' side conditions.
  Consider
  \begin{align}
    g\;x&\define
    \LET\;t\define f\;x\;\IN\;λp﹒p\;t\\
    \intertext{invoked in:}
    h\;x&\define
    \LET\;c\define g\;x\;\IN
    \;(c\;(λt﹒t))+(c\;(λt﹒(λu﹒t\times u))\;π)
  \end{align}
  The programmer would expect $h\;8$ to call~$f$ once, in the calculation of
  the temporary $t=f\;8$.
  And indeed this is what would occur in practice.
  Now consider $𝒟\;h\;8$.
  The strategy discussed above would (in the absence of memoization or
  similar heroic measures) end up calculating $f\;8$ twice, as the delayed tag
  allocation would end up splitting into two independent tag allocations with
  each independently redoing the calculation.
  This violates the constant-factor-overhead complexity guarantee of Forward
  AD, imposing, in the worst case, exponential overhead.
\end{itemize}

\section{A Second Solution: Tag Substitution}
\label{sec:second}

Another solution would be to wrap~$\bar{g}$ with tag substitution to guard
against tag collision, replacing~\eqref{eq:I} with:
\begin{align}
  \label{eq:J}
  \TANGENT\;ε₁\;\bar{g}\;y&\define\fresh\;
  (\SUBST{ε₁}{ε}\circ(\TANGENT\;ε₁)\circ\bar{g}\circ
  \SUBST{ε}{ε₁})\;y
  &
  \text{$\bar{g}$ is a function}
\end{align}
Here $\SUBST{ε₁}{ε₂}\;x$ substitutes~$ε₁$ for~$ε₂$ in~$x$.
In a language with opaque closures, tag substitution must operate on functions
by appropriate pre- and post-composition.
\begin{subequations}
  \begin{align}
  \label{eq:K1}
  \SUBST{ε₁}{ε₂}\;a&\define a&\text{$a:ℝ$}\\
  \label{eq:K2}
  \SUBST{ε₁}{ε₂}\;(a+bε₂)&\define a+bε₁\\
  \label{eq:K3}
  \SUBST{ε₁}{ε₂}\;(a+bε)&\define
  (\SUBST{ε₁}{ε₂}\;a)+(\SUBST{ε₁}{ε₂}\;b)ε&ε\not=ε₂\\
  \label{eq:K4}
  \SUBST{ε₁}{ε₂}\;\bar{g}\;y&
  \define\fresh(\SUBST{ε₂}{ε}\circ\SUBST{ε₁}{ε₂}\circ\bar{g}\circ
  \SUBST{ε}{ε₂})\;y&\text{$\bar{g}$ is a function}
  \end{align}
\end{subequations}
The intent of~\eqref{eq:K4} is to substitute~$ε₁$ for~$ε₂$ in values
closed-over in~$\bar{g}$.
An~$ε₂$ in the output of~$\bar{g}$ can result either from closed-over
values and/or input values.
We want to substitute for instances of~$ε₂$ in the output that result from
the former but not the latter.
This is accomplished by substituting a fresh tag for instances of~$ε₂$ in the
input and substituting them back at the output to preserve the extensional
behavior of~$\bar{g}$.
Equation~\eqref{eq:J} operates in a similar fashion.
The intent of~\eqref{eq:J} is to extract the coefficient of instances of~$ε₁$
in the output of~$\bar{g}$ that result from closed-over values, not input
values.
This is accomplished by substituting a fresh tag for instances of~$ε₁$ in the
input and substituting them back at the output to preserve the extensional
behavior of~$\bar{g}$.

We can see that this also resolves the bug in (\ref{eq:O1}--\ref{eq:O15}) and
accomplishes the desiderata in (\ref{eq:P1}--\ref{eq:P12}) without making two
copies of~$\Dh$.
\begin{subequations}
\begin{align}
  &\Dh\nonumber\\
  \label{eq:Q1}
  ⟹&\justification{by~\eqref{eq:Dh-definition}}\nonumber\\
  &𝒟\;s\;0\\
  \label{eq:Q2}
  ⟹&\justification{by~\eqref{eq:G}}\nonumber\\
  &\fresh\;\TANGENT\;ε\;(s\;(0+1ε))\\
  \label{eq:Q3}
  ⟹&\justification{allocate a fresh tag~$\eps₀$}\nonumber\\
  &\TANGENT\;\eps₀\;(s\;(0+1\!\!\eps₀))\\
  \label{eq:Q4}
  ⟹&\justification{by~\eqref{eq:s}}\nonumber\\
  &
  \TANGENT\;\eps₀\;(λf﹒λx﹒(f\;(x+1\!\!\eps₀)))\\
  \label{eq:Q5}
  ⟹&\justification{by~\eqref{eq:J}}\nonumber\\
  &λy﹒(\fresh\;
    (\SUBST{\eps₀\!\!}{ε}\circ(\TANGENT\;\eps₀)\circ
    (λf﹒λx﹒(f\;(x+1\!\!\eps₀)))\circ
    \SUBST{ε}{\!\!\eps₀})\;y)\\
  \hline
  &\Dh\;(\Dh\;h)\;y\nonumber\\
  \label{eq:Q6}
  ⟹&\justification{substitute~\eqref{eq:Q5} for~$\Dh$}\nonumber
\end{align}
\begin{align}
  &
  \begin{array}[t]{@{}l@{}}
    λy﹒(\fresh\;
      (\SUBST{\eps₀\!\!}{ε}\circ(\TANGENT\;\eps₀)\circ
      (λf﹒λx﹒(f\;(x+1\!\!\eps₀)))\circ
      \SUBST{ε}{\!\!\eps₀})\;y)\\
    (\begin{array}[t]{@{}l@{}}
      λy﹒(\fresh\;
        (\SUBST{\eps₀\!\!}{ε}\circ(\TANGENT\;\eps₀)\circ
        (λf﹒λx﹒(f\;(x+1\!\!\eps₀)))\circ
        \SUBST{ε}{\!\!\eps₀})\;y)\;
      h)\end{array}\\
    y
  \end{array}\\
  \label{eq:Q7}
  ⟹&\justification{beta reduce}\nonumber\\
  &
  \begin{array}[t]{@{}l@{}}
    λy﹒(\fresh\;
      (\SUBST{\eps₀\!\!}{ε}\circ(\TANGENT\;\eps₀)\circ
      (λf﹒λx﹒(f\;(x+1\!\!\eps₀)))\circ
      \SUBST{ε}{\!\!\eps₀})\;y)\\
    (\fresh\;
      (\SUBST{\eps₀\!\!}{ε}\circ(\TANGENT\;\eps₀)\circ
      (λf﹒λx﹒(f\;(x+1\!\!\eps₀)))\circ
      \SUBST{ε}{\!\!\eps₀})\;h)\\
    y
  \end{array}\\
  \label{eq:Q8}
  ⟹&\justification{beta reduce}\nonumber\\
  &
  \begin{array}[t]{@{}l@{}}
    (\fresh\;
      (\SUBST{\eps₀\!\!}{ε}\circ(\TANGENT\;\eps₀)\circ
      (λf﹒λx﹒(f\;(x+1\!\!\eps₀)))\circ
      \SUBST{ε}{\!\!\eps₀})\\
      (\fresh\;
        (\SUBST{\eps₀\!\!}{ε}\circ(\TANGENT\;\eps₀)\circ
        (λf﹒λx﹒(f\;(x+1\!\!\eps₀)))\circ
        \SUBST{ε}{\!\!\eps₀})\;h))\\
    y
  \end{array}\\
  \label{eq:Q9}
  ⟹&\justification{allocate a fresh tag~$\eps₁$}\nonumber\\
  &
  \begin{array}[t]{@{}l@{}}
    (\begin{array}[t]{@{}l@{}}
      (\SUBST{\eps₀\!\!}{\!\!\eps₁}\circ(\TANGENT\;\eps₀)\circ
      (λf﹒λx﹒(f\;(x+1\!\!\eps₀)))\circ
      \SUBST{\eps₁\!\!}{\!\!\eps₀})\\
      (\fresh\;
        (\SUBST{\eps₀\!\!}{ε}\circ(\TANGENT\;\eps₀)\circ
        (λf﹒λx﹒(f\;(x+1\!\!\eps₀)))\circ
        \SUBST{ε}{\!\!\eps₀})\;h))\end{array}\\
    y
  \end{array}\\
  \label{eq:Q10}
  ⟹&\justification{allocate a fresh tag~$\eps₂$}\nonumber\\
  &
  \begin{array}[t]{@{}l@{}}
    (\begin{array}[t]{@{}l@{}}
      (\SUBST{\eps₀\!\!}{\!\!\eps₁}\circ(\TANGENT\;\eps₀)\circ
      (λf﹒λx﹒(f\;(x+1\!\!\eps₀)))\circ
      \SUBST{\eps₁\!\!}{\!\!\eps₀})\\
      (\begin{array}[t]{@{}l@{}}
        (\SUBST{\eps₀\!\!}{\!\!\eps₂}\circ(\TANGENT\;\eps₀)\circ
        (λf﹒λx﹒(f\;(x+1\!\!\eps₀)))\circ
        \SUBST{\eps₂\!\!}{\!\!\eps₀})\;h))\end{array}\\\end{array}\\
    y
  \end{array}\\
  \label{eq:Q11}
  ⟹&\hspace*{36pt}\{\text{substitute~$\eps₂$ for~$\eps₀$, which leaves~$h$
  unchanged since it can't close over}\nonumber\\
  &\hspace*{36pt}\;\;\text{the freshly allocated tags}\}\nonumber\\
  &
  \begin{array}[t]{@{}l@{}}
    (\begin{array}[t]{@{}l@{}}
      (\SUBST{\eps₀\!\!}{\!\!\eps₁}\circ(\TANGENT\;\eps₀)\circ
      (λf﹒λx﹒(f\;(x+1\!\!\eps₀)))\circ
      \SUBST{\eps₁\!\!}{\!\!\eps₀})\\
      (\begin{array}[t]{@{}l@{}}
        (\SUBST{\eps₀\!\!}{\!\!\eps₂}\circ(\TANGENT\;\eps₀)\circ
        (λf﹒λx﹒(f\;(x+1\!\!\eps₀))))\;h))
      \end{array}\\\end{array}\\
    y
  \end{array}\\
  \label{eq:Q12}
  ⟹&\justification{beta reduce and postcompose}\nonumber\\
  &
  \begin{array}[t]{@{}l@{}}
    (\begin{array}[t]{@{}l@{}}
      (\SUBST{\eps₀\!\!}{\!\!\eps₁}\circ(\TANGENT\;\eps₀)\circ
      (λf﹒λx﹒(f\;(x+1\!\!\eps₀)))\circ
      \SUBST{\eps₁\!\!}{\!\!\eps₀})\\
      (\begin{array}[t]{@{}l@{}}
        λx﹒
        (\SUBST{\eps₀\!\!}{\!\!\eps₂}\;
        (\TANGENT\;\eps₀\;(h\;(x+1\!\!\eps₀))))))
      \end{array}\\\end{array}\\
    y
  \end{array}\\
  \label{eq:Q13}
  ⟹&\justification{substitute~$\eps₁$ for~$\eps₀$}\nonumber\\
  &
  \begin{array}[t]{@{}l@{}}
    (\begin{array}[t]{@{}l@{}}
      (\SUBST{\eps₀\!\!}{\!\!\eps₁}\circ(\TANGENT\;\eps₀)\circ
      (λf﹒λx﹒(f\;(x+1\!\!\eps₀))))\\
      (\begin{array}[t]{@{}l@{}}
        λx﹒
        (\SUBST{\eps₁\!\!}{\!\!\eps₂}\;
        (\TANGENT\;\eps₁\;(h\;(x+1\!\!\eps₁))))))
      \end{array}\\\end{array}\\
    y
  \end{array}\\
  \label{eq:Q14}
  ⟹&\justification{beta reduce and postcompose}\nonumber\\
  &(λx﹒(\SUBST{\eps₀\!\!}{\!\!\eps₁}\;(\TANGENT\;\eps₀\;
    ((λx\;.
      (\;\SUBST{\eps₁\!\!}{\!\!\eps₂}\;
      (\TANGENT\;\eps₁\;(h\;(x+1\!\!\eps₁)))))\;
      (x+1\!\!\eps₀)))))\;
    y\\
  \label{eq:Q15}
  ⟹&\justification{beta reduce}\nonumber\\
  &
  \SUBST{\eps₀\!\!}{\!\!\eps₁}\;(\TANGENT\;\eps₀\;((λx﹒
  (\SUBST{\eps₁\!\!}{\!\!\eps₂}\;(\TANGENT\;\eps₁\;(h\;(x+1\!\!\eps₁)))))\;
  (y+1\!\!\eps₀)))\\
  \label{eq:Q16}
  ⟹&\justification{beta reduce}\nonumber\\
  &
  \SUBST{\eps₀\!\!}{\!\!\eps₁}\;
  (\TANGENT\;\eps₀\;(\SUBST{\eps₁\!\!}{\!\!\eps₂}\;
  (\TANGENT\;\eps₁\;(h\;((y+1\!\!\eps₀)+1\!\!\eps₁)))))\\
  \label{eq:Q17}
  ⟹&\justification{apply~$h$ to a dual number}\nonumber
\end{align}
\begin{align}
  &
  \SUBST{\eps₀\!\!}{\!\!\eps₁}\;
  (\TANGENT\;\eps₀\;(\SUBST{\eps₁\!\!}{\!\!\eps₂}\;
  (\TANGENT\;\eps₁\;(h(y+1\!\!\eps₀)+h'(y+1\!\!\eps₀)\!\!\eps₁))))\\
  \label{eq:Q18}
  ⟹&\justification{apply~$h$ to a dual number}\nonumber\\
  &
  \SUBST{\eps₀\!\!}{\!\!\eps₁}\;
  (\TANGENT\;\eps₀\;(\SUBST{\eps₁\!\!}{\!\!\eps₂}\;
  (\TANGENT\;\eps₁\;((h(y)+h'(y)\!\!\eps₀)+h'(y+1\!\!\eps₀)\!\!\eps₁))))\\
  \label{eq:Q19}
  ⟹&\justification{apply~$h$ to a dual number}\nonumber\\
  &
  \SUBST{\eps₀\!\!}{\!\!\eps₁}\;
  (\TANGENT\;\eps₀\;(\SUBST{\eps₁\!\!}{\!\!\eps₂}\;
  (\TANGENT\;\eps₁\;
  ((h(y)+h'(y)\!\!\eps₀)+(h'(y)+h''(y)\!\!\eps₀)\!\!\eps₁))))\\
  \label{eq:Q20}
  ⟹&\justification{by~\eqref{eq:E}}\nonumber\\
  &
  \SUBST{\eps₀\!\!}{\!\!\eps₁}\;
  (\TANGENT\;\eps₀\;(\SUBST{\eps₁\!\!}{\!\!\eps₂}\;
  (h'(y)+h''(y)\!\!\eps₀)))\\
  \label{eq:Q21}
  ⟹&\justification{substitute~$\eps₁$ for~$\eps₂$}\nonumber\\
  &
  \SUBST{\eps₀\!\!}{\!\!\eps₁}\;(\TANGENT\;\eps₀\;(h'(y)+h''(y)\!\!\eps₀))\\
  \label{eq:Q22}
  ⟹&\justification{by~\eqref{eq:E}}\nonumber\\
  &
  \SUBST{\eps₀\!\!}{\!\!\eps₁}\;h''(y)\\
  \label{eq:Q23}
  ⟹&\justification{substitute~$\eps₀$ for~$\eps₁$}\nonumber\\
  &h''(y)
\end{align}
\end{subequations}
Steps~\eqref{eq:Q11} and~\eqref{eq:Q13} are abbreviated as they really
use~\eqref{eq:K4}.
Here, the tag substitution in~\eqref{eq:Q13} allows~\eqref{eq:Q17} to correct
the mistake in~\eqref{eq:O12}, just like \eqref{eq:P18}.
The implementation in Appendix~\ref{app:implementation} illustrates that this
resolves the bug when setting \lstinline{*tag-substitution?*} to
\lstinline{#t} to use the definition in~\eqref{eq:J} instead of that in
\eqref{eq:I}.

\subsection{Issues with Tag Substitution}
\label{sec:second-issues}

This solution presents several problems, when implemented as user code in a
pure language.
In the presence of aggregates, unless care is taken, the computational burden
of tag substitution can violate the complexity guarantees of Forward AD.\@
The call to $\TANGENT$ in step~\ref{step:C} might take longer than unit time as
tag substitution must potentially traverse an aggregate of arbitrary size.
When that aggregate shares substructure, a careless implementation might
traverse such shared substructure multiple times, leading to potential
exponential growth in time complexity.
Moreover, a careless implementation might copy shared substructure multiple
times, leading to potential exponential growth in space complexity.
Laziness, memoization, and hash-consing might solve this, but it can be tricky
to employ such in a fashion that preserves the requisite time and space
complexity guarantees of Forward AD, particularly in a pure or multithreaded
context.

We are unsure, however, that laziness, memoization, and hash-consing
completely eliminate the problem.
First, some languages like \Python\ and \Scheme\ lack the requisite pervasive
default laziness.
Failure to explicitly code the correct portions of user code as lazy in an
eager language can break the complexity guarantees in subtle ways.
But there are subtle issues even in languages like \Haskell\ with the requisite
pervasive default laziness, and even when laziness is correctly introduced
manually in eager languages.
One is that memoization and hash-consing implicitly involve a notion of
equality.
But it is not clear what notion of equality to use, especially with `gensym' and
potential alpha equivalence.
One might need \texttt{eq?}, \ie\ pointer or intensional equivalence, rather
than \texttt{equal?}, \ie\ structural or extensional equivalence, and all of
the impurity that this introduces.
Further, memoization and hash-consing might themselves be a source of a new
kind of perturbation confusion if tags can persist.
One would then need to substitute the memoized tags or the hash-cons cache.
Beyond this, memoization and hash-consing could break space complexity
guarantees unless the cache were flushed.
It is not clear when/where to flush the cache, and even whether there is a
consistent place to do so.
There might be inconsistent competing concerns.
Finally, many systems don't provide the requisite hooks to do all of this.
One would need weak pointers and finalization.
All of this deserves further investigation.

The above difficulties only arise when implementing tag substitution as user
code in a pure language.
The opacity of closures necessitates implementing tag substitution on
functions via pre- and post-composition \eqref{eq:K4}.
The complexity guarantees of Forward AD could be maintained if the
substitution mechanism $\SUBST{ε₁}{ε₂}\;x$ were implemented so that it
\begin{compactenum}[a)]
\item did not traverse shared substructure multiple times,
\item copied shared substructure during renaming in a fashion that preserved
  structure sharing, and
\item could apply to closures, by accessing, copying, renaming, and reclosing
  around the environments inside closures, without resorting to pre- and
  post-composition.
\end{compactenum}
This could be accomplished either by including the $\SUBST{ε₁}{ε₂}\;x$
mechanism as a primitive in the implementation, or by providing other
lower-level primitives out of which it could be fashioned.
One such mechanism is \texttt{map-closure}, the ability to reflectively access
and modify closure environments \citep{siskind-pearlmutter-popl-2007}.

\section{Differential Geometry and the Push Forward Operator}
\label{sec:DG}

The definition~\eqref{eq:partial} only extends~$𝔻$, and the mechanisms
of~§\ref{sec:first} and~§\ref{sec:second} only extend~$𝒟$, to higher-order
functions $ℝ→α$ whose ranges are functions.
Differential geometry provides the framework for extending~$𝔻$ to functions
$α₁→α₂$ whose domains too are functions.

Differential geometry concerns itself with differentiable mappings between
manifolds, where intuitively a manifold is a surface along which points can
move smoothly, like the surface of a sphere or the space of $n\times n$
rotation matrices.
Given a point~$x$, called a \defoccur{primal} (value), on a manifold~$α$,
we can consider infinitesimal perturbations of~$x$.
The space of such perturbations is a vector space called a \defoccur{tangent
  space}, denoted by $\TANGENTSPACE{x}α$.
This is a dependent type, dependent on the primal~$x$.
A particular perturbation, an element~$\tangent{x}$ of the tangent space, is
called a \defoccur{tangent} (value).
A pair~$(x,\tangent{x})$ of a primal and tangent value is called a
\defoccur{bundle} (value), which are members of a bundle space
$Tα=\sum_{x:α}\{x\}\times\TANGENTSPACE{x}α$.
Bundles generalize the notion of dual numbers.
So if~$x$ has type!$α$, for some~$α$, the tangent~$\tangent{x}$ has
type~$\TANGENTSPACE{x}α$, and they can be bundled together as
$(x+\tangent{x}\!\!\eps)$ which has type~$Tα$.

The machinery of differential geometry defines $\TANGENTSPACE{x}α$ for
various manifolds and spaces~$α$.
For function spaces $α→\beta$, where
$f$ is of type $α→\beta$, $\TANGENTSPACE{f}(α→\beta) =
(a:α)→\TANGENTSPACE{f(a)}\beta$ and\\%
$T(α→\beta)=α→ T\beta$.
The function $\Bundle\;(x:α)\;(\tangent{x}:\TANGENTSPACE{x}α)\mapsto
(x,\tangent{x}): Tα$
constructs a bundle from a primal and a tangent, and the function
$\Tangent\;(x,\tangent{x}):Tα\mapsto\tangent{x}:\TANGENTSPACE{x}α$
extracts a tangent from a bundle.
Differential geometry provides a \defoccur{push forward} operator that
generalizes the notion of a univariate derivative from functions~$f$ of type
$ℝ→ℝ$ to functions~$f$ of type $α→\beta$.
\begin{equation}
  \pf:(α→\beta)→(Tα→ T\beta)
  \label{eq:push-forward-signature}
\end{equation}
This augments the original mapping $(a:α)→\beta$ to also
\emph{linearly} map a tangent $\TANGENTSPACE{a}α$ of the input~$a$  to a
tangent $\TANGENTSPACE{f(a)}\beta$ of the output~$f(a)$.

Here we sketch how to materialize differential geometry as program constructs
to generalize~$𝔻$ to functions $α₁→α₂$ whose domains (and ranges) are
functions.
A full treatment is left for future work.
We first note that:
\begin{equation}
  \label{eq:DGa}
  𝔻\;f\;x=\Tangent\;(\pf\;f\;(\Bundle\;x\;1))
\end{equation}
This only applies when $x:ℝ$ because of the constant~$1$.
We can generalize this to a directional derivative:
\begin{equation}
  \label{eq:DGb}
  \JJ\;f\;x\;\tangent{x}=\Tangent\;(\pf\;f\;(\Bundle\;x\;\tangent{x}))
\end{equation}
This further generalizes to~$x$ of any type.
With this, $𝔻$~becomes a special case of~$\JJ$:
\begin{align}
  \label{eq:H}
  𝔻\;f\;x=\JJ\;f\;x\;1
\end{align}
To materialize~$\JJ$ in~\eqref{eq:DGb}, we need to materialize $\Tangent$,
$\pf$, and $\Bundle$.
The definition of $\TANGENT$ in (\ref{eq:D}--\ref{eq:F}, \ref{eq:I})
materializes $\Tangent$ with the first solution, Eta
Expansion~(§\ref{sec:first}), while that in (\ref{eq:D}--\ref{eq:F},
\ref{eq:J}) does so with the second solution, Tag
Substitution~(§\ref{sec:second}).
The nonstandard interpretation of the arithmetic basis functions sketched in
(\ref{eq:plus}--\ref{eq:times}) materializes $\pf$ by lifting a computation on
real numbers to a computation on dual numbers.
All that remains is to materialize $\Bundle$.
So far, we have been simply writing this as step~\ref{step:B}, a map from~$a$
to $a+1\!\!\eps$ or a map from $x$ to $x+1ε$ in~\eqref{eq:G}.
This only works for numbers, not functions.
With the framework of the first solution, Eta Expansion~(§\ref{sec:first}),
we can extend this to functions:
\begin{subequations}
\begin{align}
  \label{eq:D1}
  \BUNDLE\;ε\;x\;\tangent{x}&\define x+\tangent{x}ε
  &\text{$x$ and $\tangent{x}$ are not functions}\\
  \label{eq:I1}
  \BUNDLE\;ε\;f\;\tangent{f}\;y&\define
  \BUNDLE\;ε\;(f\;y)\;(\tangent{f}\;y)
  &\text{$f$ and $\tangent{f}$ are functions}
\end{align}
\end{subequations}
Recalling footnote~\ref{foot:postcomposition} on
page~\pageref{foot:postcomposition}, the postcomposition in~\eqref{eq:I1} is
analogous to that in~\eqref{eq:I}.
With the framework of the second solution, Tag Substitution~(§\ref{sec:second}),
we would need the alternative:
\begin{align}
  \label{eq:J1}
  \BUNDLE\;ε₁\;f\;\tangent{f}\;y&\define
  \begin{array}[t]{@{}l@{}}
    \mathbf{fresh}\;ε\\
    \mathbf{in}\;
    \SUBST{ε₁}{ε}\;(\BUNDLE\;ε₁\;
    (f\;(\SUBST{ε}{ε₁}\;y))\;
    (\tangent{f}\;(\SUBST{ε}{ε₁}\;y)))\hspace*{-80pt}
  \end{array}
  &\text{$f$ and $\tangent{f}$ are functions}
\end{align}
to~\eqref{eq:I1}.
The additional tag substitution in~\eqref{eq:J1} is analogous to that
in~\eqref{eq:J}.
With this, we can now materialize~$\JJ$ in the framework of the first
solution, Eta Expansion~(§\ref{sec:first}):
\begin{subequations}
\begin{align}
  \label{eq:U1}
  \J\;f\;x\;\tangent{x}&\defineλy﹒(\J\;(λx﹒(f\;x\;y))\;x\;\tangent{x})&
  \text{$(f\;x)$ is a function}\\
  \label{eq:V1}
  \J\;f\;x\;\tangent{x}&\define
  \fresh\;\TANGENT\;ε\;(f\;(\BUNDLE\;ε\;x\;\tangent{x}))&
  \text{$(f\;x)$ is not a function}
\end{align}
\end{subequations}
which is analogous to (\ref{eq:U}--\ref{eq:V}), and in the framework of the
second solution, Tag Substitution~(§\ref{sec:second}):
\begin{align}
  \label{eq:G1}
  \J\;f\;x\;\tangent{x}\define
  \fresh\;\TANGENT\;ε\;(f\;(\BUNDLE\;ε\;x\;\tangent{x}))
\end{align}
which is analogous to~\eqref{eq:G}.
With this, $𝒟$~becomes a special case of~$\J$:
\begin{align}
  \label{eq:H1}
  𝒟\;f\;x\define\J\;f\;x\;1
\end{align}
The implementation in Appendix~\ref{app:implementation} illustrates this
when setting \lstinline{*section9?*} to \lstinline{#t} to use~\eqref{eq:H1}
instead of either (\ref{eq:U}--\ref{eq:V}) or~\eqref{eq:G} .
Moreover, the implementation in Appendix~\ref{app:implementation} illustrates
that:
\begin{subequations}
\begin{align}
  \MAPPAIR\;f\;l&\define(f\;(\FIRST\;l)),(f\;(\SECOND\;l))\\
  \SQUARE\;x&\define x\times x\\
  \J\;\MAPPAIR\;\SQUARE\;(𝒟\;\SQUARE)\;(5,10)&⟹(10,20)
\end{align}
\end{subequations}

There is a crucial difference, however, between $\Bundle$ and $\Tangent$ and
the corresponding materializations $\BUNDLE$ and $\TANGENT$.
The former do not take~$ε$ as an argument.
This allows them to be used as distinct notational entities.
In contrast, $\BUNDLE$ and $\TANGENT$ must take the \emph{same}~$ε$ as an
argument, this tag \emph{must} be fresh, and it should not be used anywhere
else.
Thus it should not escape, except in ways that are protected by Tag
Substitution.
This motivates creation of the~$\J$ construct.
There is no corresponding standard~$\JJ$ construct in differential geometry;
we created it just to describe the intended meaning of~$\J$.

This generalization still suffers from the poor complexity properties
in~§\ref{sec:first-issues} and~§\ref{sec:second-issues}.
We don't know how to provide a materialization of differential geometry or a
program construct that can take derivatives of higher-order functions whose
domains and/or ranges include (higher order) functions in a fashion that
exhibits the complexity guarantees of Forward AD.
Moreover, we don't even know whether it is possible.

\section{Conclusion}

Classical AD systems, such as \Adifor\ \citep{Bischof1992AGD},
\Tapenade\ \citep{Hascoet2004TUG}, and
\FADBADplusplus\ \citep{Bendtsen1996FaF}, were implemented for first-order
languages like \Fortran, \Clang, and \Cplusplus.
This made it difficult to formulate situations like~\eqref{eq:Z} where
the kind of perturbation confusion reported by \citet{SiskindPearlmutter2005a}
can arise.
Thus classical AD systems did not implement the tagging mechanisms reported by
\citet{pearlmutter-siskind-popl-2007} and \citet{SiskindPearlmutter2008a}.
Moreover, such classical AD systems do not expose a derivative-taking operator
as a higher-order function, let alone one that can take derivatives of
higher-order functions.
In these systems, it is difficult to formulate the bug in~§\ref{sec:bug}.

Note that the difficulty arises from the nature of the language whose code is
differentiated and not the fact that many classical systems like \Adifor\ and
\Tapenade\ expose AD to the user via a source-code transformation implemented
via a preprocessor rather than a higher-order function.
Conceptually, both a higher-order function and a preprocessor applying a
transformation to source code map functions to functions.
Thus while one might write:
\begin{align}
  \begin{array}[t]{@{}l@{}}
    \LET\;f'\define𝒟\;f\\
    \IN\;… f'(x)…
  \end{array}
\end{align}
in a system that exposes AD to the user with an interface as a higher-order
function~$𝒟$, one would accomplish essentially the same thing in a system
that exposes AD to the user with a preprocessor that implements a source-code
transformation by having the preprocessor compute the let binding
$f'\define𝒟\;f$.
The issue presented in this manuscript would arise even in a framework that
exposes AD to the user with a preprocessor that implements a source-code
transformation if one would write
\begin{align}
  \begin{array}[t]{@{}l@{}}
    \LET\;s'\define𝒟\;s\\
    \IN\;\begin{array}[t]{@{}l@{}}
    \LET\;\Dh\define s'\;0\\
    \IN\;\Dh\;(\Dh\;h)\;y
    \end{array}
  \end{array}
\end{align}
and have the preprocessor compute the let binding $s'\define𝒟\;s$.
The difficulty in formulating the issue presented in this manuscript follows
from the fact that classical languages like \Fortran, \Clang, and
\Cplusplus\ lack the capacity for higher-order functions (closures) needed to
perform the let binding $\Dh\define s'\;0$, not from any aspect of the
difference between exposing AD via an interface via a higher-order function
\vs\ a preprocessor that implements a source-code transformation.
Indeed, the issue described here would manifest in a system that exposed
AD via a preprocessor that implements a source-code transformation in a
language such as \Python\ that supports the requisite closures and
higher-order functions (\eg\ \Myia, \citealp{breuleux2017automatic} and
\TANGENTwilt, \citealp{van2018tangent}).

Recent AD systems, such as \Myia, \TANGENTwilt, and those in
Footnote~\ref{foot:systemlist} on page~\pageref{foot:systemlist}, as well as
the \HaskellAD\ package available on Cabal \citep{Kmett2010}, the ``Beautiful
Differentiation'' system \citep{elliott2009beautiful}, and the ``Compiling to
Categories'' system \citep{elliott2017compiling}, have been implemented for
higher-order languages like \Scheme, \ML, \Haskell, \Fsharp, \Python, \Lua,
and \Julia.
One by one, many of these systems have come to discover the the kind of
perturbation confusion reported by \citet{SiskindPearlmutter2005a} and have
come to implement the tagging mechanisms reported by
\citet{pearlmutter-siskind-popl-2007} and \citet{SiskindPearlmutter2008a}.
Moreover, all these recent systems expose a derivative-taking operator as a
higher-order function.
However, except for \SCMUTILS, none supported taking derivatives of higher-order
functions.

Prior to its 30-Aug-2011 release, \SCMUTILS, the only Forward AD system that
supported taking derivatives of higher-order functions, employed the mechanism
of (\ref{eq:D}--\ref{eq:I}) and exhibited the bug
in~§\ref{sec:bug}.
An attempt was made to fix this bug in the 30-Aug-2011 release of \SCMUTILS,
using the second solution, Tag Substitution, discussed in~§\ref{sec:second},
in response to an early version of this manuscript.
\SCMUTILS\ was patched to include code that is similar to, but not identical
to, \eqref{eq:J} and (\ref{eq:K1}--\ref{eq:K4}).
Crucially, it allocates a fresh tag in its implementation of~\eqref{eq:J} but
not in its implementation of~\eqref{eq:K4}; its implementation
of~\eqref{eq:K4} being
\begin{align}
  \label{eq:K4a}
  \SUBST{ε₁}{ε₂}\;\bar{g}&\define
  \SUBST{ε₂}{ε₁}\circ\bar{g}\circ\SUBST{ε₁}{ε₂}.
  &\text{$\bar{g}$ is a function}
\end{align}
This, however, is incorrect, as illustrated by the following variant of the
bug in §\ref{sec:bug}:
\begin{align}
  \label{eq:bugA}
  v\;u\;f₁\;f₂\;x&\define f₁\;f₂\;(x + u)\\
  \label{eq:bugB}
  i\;x&\define x
\end{align}
Variants of (\ref{eq:L1}--\ref{eq:L3}) show that
$𝔻\;v\;0\;(𝔻\;v\;0\;i)\;h\;y=h''(y)$.
The 27-Aug-2016 release, the current release at the time of writing,
however, yields $𝒟\;v\;0\;(𝒟\;v\;0\;i)\;h\;y\reducesto0$.
Both solutions presented here yield the correct result.

In 2019, the authors reached out to Gerald Jay Sussman, one of the authors of
\SCMUTILS, to help fix \SCMUTILS.
He asked whether we could produce an example that illustrated the necessity of
performing substitution on functions \eqref{eq:K4} and why an alternate
\begin{align}
  \label{eq:K4b}
  \SUBST{ε₁}{ε₂}\;\bar{g}&\define\bar{g}&\text{$\bar{g}$ is a function}
\end{align}
that did not perform substitution on functions wouldn't suffice.
A variant of (\ref{eq:s}, \ref{eq:Dh-definition}) that wraps and unwraps
arguments and results in Church-encoded boxes illustrates the necessity of
\eqref{eq:K4}.
\begin{subequations}
\begin{align}
  \textsc{box}&:ℝ→\Box\;ℝ\nonumber\\
  \textsc{box}\;x\;m&\define m\;x\\[2ex]
  \textsc{unbox}&:\Box\;ℝ→ℝ\nonumber\\
  \textsc{unbox}\;x&\define x\;(λx﹒x)\\[2ex]
  \textsc{wrap}&:(ℝ→ℝ)→(\Box\;ℝ→\Box\;ℝ)\nonumber\\
  \textsc{wrap}\;f\;x&\define\textsc{box}\;(f\;(\textsc{unbox}\;x))\\[2ex]
  \textsc{unwrap}&:(\Box\;ℝ→\Box\;ℝ)→(ℝ→ℝ)\nonumber\\
  \textsc{unwrap}\;f\;x&\define\textsc{unbox}\;(f\;(\textsc{box}\;x))\\[2ex]
  \textsc{wrapTwo}&:((ℝ→ℝ)→(ℝ→ℝ))→((\Box\;ℝ→\Box\;ℝ)→(\Box\;ℝ→\Box\;ℝ))
  \nonumber\\
  \textsc{wrapTwo}\;f\;g\;x&\define
  \textsc{box}\;((f\;(\textsc{unwrap}\;g))\;(\textsc{unbox}\;x))\\[2ex]
  \textsc{wrapTwoResult}&:\nonumber\\
  &\hspace*{-40pt}(ℝ→((ℝ→ℝ)→(ℝ→ℝ)))→(ℝ→((\Box\;ℝ→\Box\;ℝ)→(\Box\;ℝ→\Box\;ℝ)))
  \nonumber\\
  \textsc{wrapTwoResult}\;f\;x&\define\textsc{wrapTwo}\;\;(f\;x)\\[2ex]
  \textsc{wrapped}\Dh&\define\D\;(\textsc{wrapTwoResult}\;s)\;0
\end{align}
The same analysis as (\ref{eq:L1}--\ref{eq:L3}) shows that:
\begin{align}
  \textsc{unwrap}\;
  (𝔻\;(\textsc{wrapTwoResult}\;s)\;0\;(𝔻\;(\textsc{wrapTwoResult}\;s)\;0\;(\textsc{wrap}\;h)))
  =h''
\end{align}
While
\begin{align}
  \textsc{unwrap}\;
  (𝒟\;(\textsc{wrapTwoResult}\;s)\;0\;(𝒟\;(\textsc{wrapTwoResult}\;s)\;0\;(\textsc{wrap}\;h)))
  =h''
\end{align}
with both \eqref{eq:K4} and \eqref{eq:K4b}, with \eqref{eq:K4},
\begin{align}
  \label{eq:bugD}
  \textsc{unwrap}\;(\textsc{wrapped}\Dh\;(\textsc{wrapped}\Dh\;(\textsc{wrap}\;h)))=h''
\end{align}
but with \eqref{eq:K4b},
\begin{align}
  \label{eq:bugC}
  \textsc{unwrap}\;(\textsc{wrapped}\Dh\;(\textsc{wrapped}\Dh\;(\textsc{wrap}\;h)))\not=h''
\end{align}
\end{subequations}
The authors of \SCMUTILS\ are in the process of fixing it again in response to
this updated manuscript.
The tenacity of this bug illustrates its subtlety and cries out for a proof of
correctness.

Practically all systems that expose a derivative-taking operator as a
higher-order function generalize that operator to take gradients and
Jacobians of functions whose domains and/or ranges are aggregates, and most
have come to implement tagging.
The current forefront of deep learning research often involves nested
application of AD and application of AD to higher-order functions
\citep{Raissi-2018a, chen-ruanova-etal-2018a, maclaurin2015gradient,
  Andrychowicz-etal-2016a, Salman-etal-2018a}.
This work often combines building custom frameworks to support the particular
derivatives of interest, and performing transformations (closure conversion or
even full AD transforms) manually.
Under the pressure of machine learning programmers' desire for nesting and for
derivatives of higher-order functions, it is reasonable to speculate that
many, if not most, of the above systems will attempt to support these usage
patterns.
We hope that the awareness provided by this manuscript will help such efforts
avoid this particular subtle bug.

Without formal proofs, we cannot really be sure whether the first solution,
Eta Expansion (\ref{eq:D}--\ref{eq:F}, \ref{eq:U}, \ref{eq:V}), or the second
solution, Tag Substitution (\ref{eq:D}--\ref{eq:G}, \ref{eq:J}), correctly
implements the specification in~\eqref{eq:partial}.
We cannot even be sure that (\ref{eq:D}--\ref{eq:G}) correctly
implement the specification in~\eqref{eq:lim}.
These are tricky due to subtleties like nondifferentiability, nontermination,
and the difference between function intensions and extensions pointed out by
\citet[footnote~1]{SiskindPearlmutter2008a}.
\citet{Ehrhard-Regnier-2003a}, \citet{manzyuk-mfps2012,
manzyuk-tan-bun-in-diff-cats}, \citet{AD2016e}, and
\citet{plotkin-2018popl} present promising work in this direction.
Given these sorts of subtle bugs, and the growing interest in---and
economic and societal importance of---complicated software systems
driven by nested automatically calculated derivatives, it is our hope
that formal methods can bridge the gap between the Calculus and the
Lambda Calculus, allowing derivatives of interest of arbitrary
programs to be not just automatically and efficiently calculated, but
also for their correctness to be formally verified.

\section*{Acknowledgments}
We would like to thank Gerald Jay Sussman for wrestling at our side in
the search for correctness.
We also appreciate the efforts of Olivier Danvy in helping improve an
earlier version of this manuscript, and Jeremy Gibbons for helping to
improve this version.

\bibliographystyle{jfp}
\newcounter{arxiv}
\setcounter{arxiv}{1}
\ifnum\value{arxiv}=1
\bibliography{jfp2017-r4}
\fi
\ifnum\value{arxiv}=0

\fi

\appendix
\section{A Minimal Implementation}
\label{app:implementation}

The repository \url{https://github.com/qobi/amazing}, file \href{https://github.com/qobi/amazing/blob/master/implementation.ss}{implementation.ss},
also available as supplementary material, contains a minimal implementation.
It is not intended as a full practical implementation but rather has the
expository purpose of explaining the ideas presented in this manuscript.
The implementations of \lstinline{list-real->real} and
\lstinline{list-real*real->real} are similar to those by
\citet[Fig.~2]{SiskindPearlmutter2008a}.
Setting both \lstinline{*eta-expansion?*} and \lstinline{*tag-substitution?*}
to \lstinline{#f} uses the implementation of~$𝒟$ in~\eqref{eq:G},
the implementation of~$\J$ in~\eqref{eq:G1}, the implementation of
$\TANGENT$ for functions in~\eqref{eq:I}, and the implementation of
$\BUNDLE$ for functions in~\eqref{eq:I1} and illustrates the bug in
(\ref{eq:O1}--\ref{eq:O15}, \ref{eq:M}).
Setting \lstinline{*eta-expansion?*} to \lstinline{#t} implements the first
solution, Eta Expansion, from~§\ref{sec:first} and uses the implementation
of~$𝒟$ in (\ref{eq:U}--\ref{eq:V}), instead of that in~\eqref{eq:G}, and
the implementation of~$\J$ in (\ref{eq:U1}--\ref{eq:V1}), instead of that
in~\eqref{eq:G1}.
This resolves the bug and yields the correct result
(\ref{eq:N1}--\ref{eq:N26}).
Here, $𝒟$ and~$\J$ each use a single side effect to generate~$\eps$s.
Instead, setting \lstinline{*tag-substitution?*} to \lstinline{#t} implements
the second solution, Tag Substitution, from~§\ref{sec:second} and uses the
implementation of $\TANGENT$ for functions in~\eqref{eq:J}, instead of that in
\eqref{eq:I}, and the implementation of $\BUNDLE$ for functions
in~\eqref{eq:J1}, instead of that in \eqref{eq:I1}.
This resolves the bug and yields the correct result
(\ref{eq:Q1}--\ref{eq:Q23}).
Here, $𝒟$, $\J$, $\TANGENT$, $\BUNDLE$, and tag substitution for functions
each use a single side effect to generate~$\eps$s.
Setting \lstinline{*section9?*} to \lstinline{#t} implements the
generalization in~§\ref{sec:DG} and uses the implementation of $𝒟$
in~\eqref{eq:H1} instead of those in~\eqref{eq:G} or (\ref{eq:U}--\ref{eq:V}).
This works with either solution but exhibits the bug when both solutions are
disabled.
In all cases, the function whose derivative is taken is pure.
This illustrates that the bug can be addressed even when an impure mechanism
is used to generate~$ε$s.
When setting \lstinline{*tag-substitution?*} to \lstinline{#t}, setting
\lstinline{*function-substitution*} to \lstinline{equation-38}
uses~\eqref{eq:K4a} and gives the wrong result
for~(\ref{eq:bugA}, \ref{eq:bugB}),
setting \lstinline{*function-substitution*} to \lstinline{equation-41}
uses~\eqref{eq:K4b} and illustrates the bug in~\eqref{eq:bugC}, while
setting \lstinline{*function-substitution*} to \lstinline{equation-24d}
uses~\eqref{eq:K4}, gives the correct result for~(\ref{eq:bugA},
\ref{eq:bugB}), and upholds~\eqref{eq:bugD}.

\ifnum\value{arxiv}=1
\begin{figure}
  \centering
  \scalebox{0.8}{\fbox{
\begin{lstlisting}[lastline=46]
#!r6rs

(define-record-type dual-number (fields epsilon primal tangent))

(define *eta-expansion?* #f)

(define *tag-substitution?* #t)

(define *section9?* #f)

;;; One of equation-24d, equation-38, or equation-41
(define *function-substitution* 'equation-24d)

(define *epsilon* 0)

(define (generate-epsilon)
 (set! *epsilon* (+ *epsilon* 1))
 *epsilon*)

(define (subst epsilon1 epsilon2 x)
 (cond ((real? x) x)			;Equation (24a)
       ((dual-number? x)
	(if (= (dual-number-epsilon x) epsilon2)
	    ;; Equation (24b)
	    (make-dual-number
	     epsilon1 (dual-number-primal x) (dual-number-tangent x))
	    ;; Equation (24c)
	    (make-dual-number
	     (dual-number-epsilon x)
	     (subst epsilon1 epsilon2 (dual-number-primal x))
	     (subst epsilon1 epsilon2 (dual-number-tangent x)))))
       ((procedure? x)
	(lambda (y)
	 (case *function-substitution*
	  ((equation-24d)
	   ;; Equation (24d)
	   (let ((epsilon3 (generate-epsilon)))
	    (subst epsilon2 epsilon3
		   (subst epsilon1 epsilon2 (x (subst epsilon3 epsilon2 y))))))
	  ((equation-38)
	   ;; Equation (38)
	   (subst epsilon2 epsilon1 (x (subst epsilon1 epsilon2 y))))
	  ;; Equation (41)
	  ((equation-41) (x y))
	  (else (error "A")))))
       (else (error "B"))))
\end{lstlisting}}}
  \caption{The implementation, Part I}
  \label{fig:implementationI}
\end{figure}

\begin{figure}
  \centering
  \scalebox{0.8}{\fbox{
\begin{lstlisting}[firstline=48,lastline=104]
(define (prim epsilon x)
 (cond ((real? x) x)
       ((dual-number? x)
	(if (= (dual-number-epsilon x) epsilon)
	    (dual-number-primal x)
	    (make-dual-number (dual-number-epsilon x)
			      (prim epsilon (dual-number-primal x))
			      (prim epsilon (dual-number-tangent x)))))
       ((procedure? x)
	(if *tag-substitution?*
	    (lambda (y)
	     (let ((epsilon2 (generate-epsilon)))
	      (subst epsilon
		     epsilon2
		     (prim epsilon (x (subst epsilon2 epsilon y))))))
	    (lambda (y) (prim epsilon (x y)))))
       (else (error "C"))))

(define (tg epsilon x)
 (cond ((real? x) 0)			;Equation (8a)
       ((dual-number? x)
	(if (= (dual-number-epsilon x) epsilon)
	    ;; Equation (8b)
	    (dual-number-tangent x)
	    ;; Equation (8c)
	    (make-dual-number (dual-number-epsilon x)
			      (tg epsilon (dual-number-primal x))
			      (tg epsilon (dual-number-tangent x)))))
       ((procedure? x)
	(if *tag-substitution?*
	    ;; Equation (23)
	    (lambda (y)
	     (let ((epsilon2 (generate-epsilon)))
	      (subst epsilon
		     epsilon2
		     (tg epsilon (x (subst epsilon2 epsilon y))))))
	    ;; Equation (8e)
	    (lambda (y) (tg epsilon (x y)))))
       (else (error "D"))))

(define (bun epsilon x x-prime)
 (cond ((and (or (real? x) (dual-number? x))
	     (or (real? x-prime) (dual-number? x-prime)))
	;; Equation (30a)
	(make-dual-number epsilon x x-prime))
       ((and (procedure? x) (procedure? x-prime))
	(if *tag-substitution?*
	    ;; Equation (31)
	    (lambda (y)
	     (let ((epsilon2 (generate-epsilon)))
	      (subst epsilon epsilon2
		     (bun epsilon
			  (x (subst epsilon2 epsilon y))
			  (x-prime (subst epsilon2 epsilon y))))))
	    ;; Equation (30b)
	    (lambda (y) (bun epsilon (x y) (x-prime y)))))
       (else (error "E"))))
\end{lstlisting}}}
  \caption{The implementation, Part II}
  \label{fig:implementationII}
\end{figure}

\begin{figure}
  \centering
  \scalebox{0.8}{\fbox{
\begin{lstlisting}[firstline=106,lastline=164]
(define (lift-real->real f df/dx)
 (letrec ((self (lambda (x)
		 (if (dual-number? x)
		     (let ((epsilon (dual-number-epsilon x)))
		      (bun epsilon
			   (self (prim epsilon x))
			   (d* (df/dx (prim epsilon x)) (tg epsilon x))))
		     (f x)))))
  self))

(define (lift-real*real->real f df/dx1 df/dx2)
 (letrec ((self (lambda (x1 x2)
		 (if (or (dual-number? x1) (dual-number? x2))
		     (let ((epsilon (if (dual-number? x1)
					(dual-number-epsilon x1)
					(dual-number-epsilon x2))))
		      (bun epsilon
			   (self (prim epsilon x1) (prim epsilon x2))
			   (d+ (d* (df/dx1 (prim epsilon x1) (prim epsilon x2))
				   (tg epsilon x1))
			       (d* (df/dx2 (prim epsilon x1) (prim epsilon x2))
				   (tg epsilon x2)))))
		     (f x1 x2)))))
  self))

;;; Equation (5a)
(define d+ (lift-real*real->real + (lambda (x1 x2) 1) (lambda (x1 x2) 1)))

;;; Equation (5b)
(define d* (lift-real*real->real * (lambda (x1 x2) x2) (lambda (x1 x2) x1)))

(define dexp (lift-real->real exp (lambda (x) (dexp x))))

(define (j* f)
 (lambda (x x-prime)
  (if *eta-expansion?*
      (if (procedure? (f x))
	  ;; Equation (32a)
	  (lambda (y) ((j* (lambda (x) ((f x) y))) x x-prime))
	  ;; Equation (32b)
	  (let ((epsilon (generate-epsilon)))
	   (tg epsilon (f (bun epsilon x x-prime)))))
      ;; Equation (33)
      (let ((epsilon (generate-epsilon)))
       (tg epsilon (f (bun epsilon x x-prime)))))))

(define (d f)
 (lambda (x)
  (cond (*section9?* ((j* f) x 1))	;Equation (34)
	(*eta-expansion?*
	 (if (procedure? (f x))
	     ;; Equation (18a)
	     (lambda (y) ((d (lambda (x) ((f x) y))) x))
	     ;; Equation (18b)
	     (let ((epsilon (generate-epsilon)))
	      (tg epsilon (f (make-dual-number epsilon x 1))))))
	;; Equation (8d)
	(else (let ((epsilon (generate-epsilon)))
	       (tg epsilon (f (make-dual-number epsilon x 1))))))))
\end{lstlisting}}}
  \caption{The implementation, Part III}
  \label{fig:implementationIII}
\end{figure}

\begin{figure}
  \centering
  \scalebox{0.8}{\fbox{
\begin{lstlisting}[firstline=166,lastline=227]
;;; Equation (9)
;;; R->((R->R)->(R->R))
(define (s u) (lambda (f) (lambda (x) (f (d+ x u)))))

;;; Equation (11)
(define d-hat ((d s) 0))

;;; Siskind & Pearlmutter (IFL 2005) Equation (2)
(write ((d (lambda (x) (d* x ((d (lambda (y) (d+ x y))) 1)))) 1))
(newline)
;;; Siskind & Pearlmutter (HOSC 2008) pages 363-364
(write ((d (lambda (x) (d* x ((d (lambda (y) (d* x y))) 2)))) 1))
(newline)
;;; Equation (13 left) with h=exp and y=1
(write ((d-hat (d-hat dexp)) 1))
(newline)
;;; Equation (13 right) with h=exp and y=1
(write ((d (d dexp)) 1))
(newline)

;;; Equation (20a)
(define (pair a) (lambda (d) (lambda (m) ((m a) d))))

;;; Equation (20b)
(define (fst c) (c (lambda (a) (lambda (d) a))))

;;; Equation (20c)
(define (snd c) (c (lambda (a) (lambda (d) d))))

;;; Equation (20d)
(define (t u) ((pair (dexp (d* u u))) (lambda (f) (lambda (x) (f (d+ x u))))))

(write ((d (lambda (x) (dexp (d* x x)))) 1))
(newline)
;;; Equation (20e)
(write (fst ((d t) 1)))
(newline)
(let* ((p ((d t) 0))			;Equation (20f)
       ;; Equation (20h)
       (d-vec (snd p)))
 ;; Equation (20g)
 (write (fst p))
 (newline)
 ;; Equation (20i)
 (write ((d-vec (d-vec dexp)) 1))
 (newline))

(define pi (* 2 (acos 0)))

(define (f x) x)

;;; Equation (21)
(define (g x) (let ((t (f x))) (lambda (p) (p t))))

;;; Equation (22)
(define (h x)
 (let ((c (g x)))
  (d+ (c (lambda (t) t)) ((c (lambda (t) (lambda (u) (d* t u)))) pi))))

;;; Example from page 12
(write ((d h) 8))
(newline)
\end{lstlisting}}}
  \caption{The implementation, Part IV}
  \label{fig:implementationIV}
\end{figure}

\begin{figure}
  \centering
  \scalebox{0.8}{\fbox{
\begin{lstlisting}[firstline=229]
;;; Equation (35a)
(define (map-pair f) (lambda (l) ((pair (f (fst l))) (f (snd l)))))

;;; Equation (35b)
(define (sqr x) (d* x x))

;;; Equation (35c)
(let ((result (((j* map-pair) sqr (d sqr)) ((pair 5) 10))))
 (write (fst result))
 (newline)
 (write (snd result))
 (newline))

;;; Equation (39)
(define (v u) (lambda (f1) (lambda (f2) (lambda (x) ((f1 f2) (d+ x u))))))

;;; Equation (40)
(define (i x) x)

;;; Example from page 20
(write (((((d v) 0) (((d v) 0) i)) dexp) 1))
(newline)

;;; Equation (42a)
;;; R->(box R)
(define (box x) (lambda (m) (m x)))

;;; Equation (42b)
;;; (box R)->R
(define (unbox x) (x (lambda (x) x)))

;;; Equation (42c)
;;; (R->R)->((box R)->(box R))
(define (wrap f) (lambda (x) (box (f (unbox x)))))

;;; Equation (42d)
;;; ((box R)->(box R))->(R->R)
(define (unwrap f) (lambda (x) (unbox (f (box x)))))

;;; Equation (42e)
;;; ((R->R)->(R->R))->(((box R)->(box R))->((box R)->(box R)))
(define (wrap2 f) (lambda (g) (lambda (x) (box ((f (unwrap g)) (unbox x))))))

;;; Equation (42f)
;;; (R->((R->R)->(R->R)))->(R->(((box R)->(box R))->((box R)->(box R))))
(define (wrap2-result f) (lambda (x) (wrap2 (f x))))

;;; Equation (42g)
(define wrapped-d-hat ((d (wrap2-result s)) 0))

;;; Equation (42i) with h-exp and y=1
(write
 ((unwrap (((d (wrap2-result s)) 0) (((d (wrap2-result s)) 0) (wrap dexp)))) 1))
(newline)

;;; Equation (42j) with h-exp and y=1,
;;;          when *function-substitution*='equation-24d
;;;       or (42k) with h-exp and y=1,
;;;          when *function-substitution*='equation-41
(write ((unwrap (wrapped-d-hat (wrapped-d-hat (wrap dexp)))) 1))
(newline)
(exit)
\end{lstlisting}}}
  \caption{The implementation, Part V}
  \label{fig:implementationV}
\end{figure}
\fi

\end{document}